\documentclass[preprint,12pt,sort&compress]{elsarticle}

\usepackage{graphicx,amssymb,amsmath}
\usepackage[dvips]{color}

\newcommand{\np}{\mbox{\boldmath $p$}}
\newcommand{\bd}[1]{ \mbox{\boldmath $#1$}  }

\usepackage{graphicx,amssymb,amsmath,color,numcompress}
\journal{Nuclear Physics B}

\begin{document}

\begin{frontmatter}



\title{Off-shell effects in the relativistic mean field model and their role in CC (anti)neutrino scattering at MiniBooNE kinematics}

\author[label1,label2]{M.~V.~Ivanov\corref{cor}}
\cortext[cor]{Corresponding author.}
\author[label3]{R.~Gonz\'alez-Jim\'enez}
\author[label3]{J.~A.~Caballero}
\author[label4]{M.B.~Barbaro}
\author[label5]{T.~W.~Donnelly}
\author[label1]{J.M. Ud\'{\i}as}

\address[label1]{Grupo\,de\,F\'{i}sica\,Nuclear,\,Departamento\,de\,F\'{i}sica\,At\'omica,\,Molecular\,y\,Nuclear,
Facultad\,de\,Ciencias\,\,F\'{i}sicas,\,Universidad\,Complutense\,de\,Madrid, CEI~Moncloa,\,Madrid\,E-28040,\,Spain}
\address[label2]{Institute\,for\,Nuclear\,Research\,and\,Nuclear\,Energy,\,Bulgarian\,Academy\,of\,Sciences,\,Sofia\,1784,\,Bulgaria}

\address[label3]{Departamento de F\'{\i}sica At\'{o}mica, Molecular y Nuclear, Universidad de Sevilla, 41080 Sevilla, Spain}

\address[label4]{Dipartimento di Fisica, Universit\`a di Torino and INFN, Sezione di Torino, Via P. Giuria 1, 10125 Torino, Italy}

\address[label5]{Center for Theoretical Physics, Laboratory for Nuclear
  Science and Department of Physics, Massachusetts Institute of Technology,
  Cambridge, MA 02139, USA}

\begin{abstract}

The relativistic mean field (RMF) model is used to describe nucleons in the nucleus and thereby to evaluate the effects of having dynamically off-shell spinors. Compared with free, on-shell nucleons as employed in some other models, within the RMF nucleons are described by relativistic spinors with strongly enhanced lower components. In this work it is seen that for MiniBooNE kinematics, neutrino charged-current quasielastic cross sections show some
sensitivity to these off-shell effects, while for the antineutrino-nucleus case the total cross sections are seen to be essentially independent of the enhancement of the lower components. As was found to be the case when comparing the RMF results with the neutrino-nucleus data, the present impulse approximation predictions within the RMF also fall short of the MiniBooNE antineutrino-nucleus data.

\end{abstract}

\end{frontmatter}

\section{Introduction\label{sec1}}

An increased interest in neutrino interactions in the few GeV energy range has emerged from the recent cross section measurements taken at different  laboratories. In particular, the MiniBooNE data on charged-current quasielastic
(CCQE)~\cite{AguilarArevalo:2010zc,CCQE-anti-exp} and neutral-current quasielastic (NCQE)~\cite{AguilarArevalo:2010cx}
$\nu_\mu$ and ${\overline\nu}_\mu$ scattering off $^{12}$C, with mean beam energy $\langle E_\nu\rangle=$ 788 and $\langle E_{\overline\nu}\rangle=$ 665 MeV, respectively, have stimulated important discussions about the role played by both nuclear and nucleonic ingredients in the description of the reaction. To characterize CCQE (and similarly NCQE) neutrino scattering from carbon, the MiniBooNE collaboration made use of the Relativistic Fermi Gas (RFG) model in Monte Carlo simulations. Indeed, the MiniBooNE cross section is underestimated by the RFG unless the axial mass $M_A$ is significantly enlarged ($1.35$~GeV/c$^2$~\cite{AguilarArevalo:2010zc}) with respect to the world average value ($1.03$~GeV/c$^2$~\cite{Bernard:2001rs}) extracted from neutrino and antineutrino scattering data off the deuteron. However, previous data from the NOMAD collaboration~\cite{Lyubushkin:2008pe} for higher beam energies (from $3$ to $100$~GeV) are in good agreement with the standard value of the axial mass, and recent data from the MINER$\nu$A
collaboration~\cite{FT2012}, corresponding to (anti)neutrino energies from $1.5$ to $10$~GeV, are claimed to disfavor the value $M_A = 1.35$ GeV/c$^2$. Furthermore, as emerges from comparisons with electron scattering data, the RFG is too simplistic to account for the nuclear dynamics, and in particular it fails badly to reproduce the separated longitudinal and transverse response functions, which is essential to make reliable predictions for neutrino scattering, where the balance between the two channels is different from the electron scattering case. A larger axial mass within the RFG should simply be interpreted as a crude way effectively to incorporate nuclear effects.

In addition to the RFG, other more sophisticated models based on the Impulse Approximation (IA) also underpredict the CCQE cross section measured at MiniBooNE~\cite{Leitner:2006ww, Martini:2009uj, Benhar:2010nx, Ankowski:2012, Nieves:2011pp}. As well the phenomenological SuSA model described in \cite{Amaro:2004bs}, based on the superscaling function extracted from quasielastic electron scattering cross sections~\cite{BillIngo}, predicts neutrino cross
sections which are found to be lower than the MiniBooNE data~\cite{Amaro:2010sd, Amaro:2011qb, Amaro:2011aa, Amaro:2013yna}.

Different explanations have been proposed, based either on multi-nucleon knockout~\cite{Martini:2009uj, Martini:2011wp, Martini2010, Nieves:2011yp, Nieves:2013fr, Amaro:2010sd, Amaro:2011qb, Amaro:2011aa} or on particular treatments of final-state interactions through phenomenological optical potentials~\cite{Meucci:2009sup,Meucci:2011vd,Meucci:2011nc}. Although there is general agreement that multi-nucleon effects produce a significant enhancement of the cross section, at a quantitative level the theoretical uncertainties related to the description of nuclear effects are rather large, as substantially different approximations are involved in each of the above-mentioned approaches.

The accurate interpretation of present experiments depends on the understanding of all ingredients of the theory. Among them one would need to address the issue of nucleons being bound in the nuclei that form the nuclear targets and thus, necessarily, the cross sections have to be computed for off-shell nucleons. Indeed, within the IA the cross section is described as a sum of lepton-nucleon vertices, where the nucleons are bound and thus off-shell. The neutrino-nucleus reaction at intermediate energies, as is the case of the MiniBooNE experiment, would show sensitivity to off-shell effects~\cite{Jeschonnek:1998}. At intermediate or low energies, lepton-nucleon reactions are often described with models of the lepton-nucleon  interaction~\cite{Leitner:2006ww, Martini:2009uj, Martini2010, Nieves:2013fr, Nieves:2011yp} that incorporate  relativistic effects into the kinematics and in some cases also into the dynamics. In the SuSA approach with Meson Exchange Currents (MEC) of \cite{Amaro:2010sd,Amaro:2011qb,Amaro:2011aa}, while both the kinematics and the one- and two-body current operators are fully relativistic, only positive energy on-shell spinors are taken into account.

Moreover, most non-relativistically inspired approaches implicitly assume on-shell nucleons, that is, only positive energy spinors are involved in the modeling~\cite{Javi2004}. Thus they are less suitable for studying the influence of off-shell effects. Work has been done~\cite{Caballero:1997gc} in a relativistic context where off-shellness in the initial-state bound nucleons was the main focus, and led to discussions of the break-down of factorization in $(e,e'p)$ reactions. In this work, working at the mean field level using one-body effective operators ({\it i.e.,} within the IA) we study the effects of off-shellness in both initial and final states within the context of the Relativistic Mean Field (RMF) model. The RMF has been successfully employed to describe electron-nucleus and neutrino-nucleus experiments~\cite{Amaro:2011qb,NC_Martin,form2} and the opportunity presented with the availability of both neutrino
and antineutrino data can shed light on off-shell effects, as we shall illustrate in this letter.

It is worth noting that even if the RMF is a one-body model, that is, processes containing other particles in addition to the nucleon in the final state -- including multi-nucleon knockout and pion production -- are not explicitly incorporated in this formalism, the one-body contribution from multi-nucleon knockout is to some extent incorporated into the model via the self-energy of the propagating nucleon. The RMF approach at mean field level includes all types of rescattering processes (elastic and inelastic) with the remaining nucleons. Here the redistribution of the strength and multi-nucleon knockout are attributed to final-state interactions and not to explicit correlations. Notice that the RMF model provides the correct saturation properties for nuclear matter already at the mean field level~\cite{SW} stemming from the combination of the strong scalar ($S$) and vector ($V$) potentials that incorporate repulsive and attractive interactions. Further, within the RMF, initial and final nucleon wave functions are computed with the same mean field equation and potentials, thus the current computed from these spinors fulfills the continuity equation.

Within the RMF, the presence of strong $S<0$ and $V>0$ potentials in the nuclear states (mainly the final one) leads to a significant enhancement of the lower components of the four-spinors describing the relativistic nucleon wave functions according to the relationship~\cite{Javi2004,Lund,Udias99}
\begin{equation}
\psi_{down}(\np)=\frac{{\bd \sigma} \cdot \np}{E+M_N+S-V}\psi_{up}(\np) \, .
\label{psibound}
\end{equation}
That is, the nucleons are dynamically and strongly off-shell. This strong off-shellness is the main cause for the lack of exact factorization of the results, even at the IA level. Thus RMF is not factorized into a spectral function and an elementary lepton-nucleus cross sections, as it is done at times in describing these reactions~\cite{Benhar:2010nx, Ankowski:2012}. Factorization break-down is however not very strong, as the results of the SuSA approach (that obviously is a factorized scheme) do not depart much from the RMF  predictions~\cite{Amaro:2006pr, Amaro:2004bs, Amaro:2011qb, Amaro:2011aa, Amaro:2010sd}.

In order to assess the influence of this strong off-shellness, also denoted in the past as spinor distortion~\cite{Kelly}, the fully relativistic results can be compared with the effective momentum approach (EMA)~\cite{Kelly, Lund, Udias, Javi2004}. Within EMA, the spinors are put exactly on the mass shell, by enforcing the same relationship between upper and lower components as for free spinors. EMA spinors lack the dynamical enhancement of the lower components due to the presence of strong potentials. Lacking this spinor distortion, the EMA results should lie closer to the so-called factorized approach~\cite{Lund,Javi2004}. The comparison between EMA and RMF is interesting because it allows one to estimate to what extent dynamical off-shell effects may affect neutrino-nucleus observables such as those involved in the analysis of MiniBooNE data --- and this in the context of a model that has been validated against inclusive electron scattering data in the quasielastic region at intermediate energies.

\section{Results and discussion}\label{sec3}

Before entering into a detailed study of results, it is helpful to keep in mind the following general properties. For CC inclusive neutrino-nucleus scattering, only the $L$, $T$ and $T'$ contributions to the cross section survive~\cite{form1,form2,form3,Alb97,Chiara03,Martinez:2005xe,JA06,Caballero:2005sj,Jin92}:
\begin{eqnarray}
\frac{d^2\sigma}{d\varepsilon_f d\cos\theta } =
\frac{d^2\sigma_L}{d\varepsilon_f d\cos\theta} +
\frac{d^2\sigma_T}{d\varepsilon_f d\cos\theta} +
h  \frac{d^2\sigma_{T'}}{d\varepsilon_f d\cos\theta} 
\label{eq:cross1}
\end{eqnarray}
where $h$ denotes the helicity of the incident lepton ($h = -1$ for neutrinos and $h = +1$ for antineutrinos), $\varepsilon_f$ and $\theta$ represent the energy and  scattering angle of the outgoing lepton. We describe the bound nucleon states as self-consistent Dirac-Hartree solutions, derived within a relativistic mean-field approach using a Lagrangian containing $\sigma$, $\omega$ and $\rho$ mesons~\cite{boundwf,SW}. The electroweak current operators are  the same as in recent work~\cite{Amaro:2004bs,JA06,AM05} and in a model-dependent way account for some aspects of
off-shellness, namely ``kinematical'' off-shellness rather than our focus in the present work which is ``dynamical'' off-shellness stemming from the bound and continuum nucleon's being off-shell with non-trivial lower components (see~\cite{Caballero:1993} for more discussion of kinematical off-shellness).

Based on the use of the CC2 current (considered in this work) the $L$ contribution is rather insensitive to off-shell effects, which can be traced back to the fact that within the RMF the matrix elements of the CC2 charge current fulfill the continuity equation already at the one-body level. Actually, within the RMF (and also under some other more general conditions,  see~\cite{Javi2004, Caballero:1997gc, Caballero:1998ip}), the $L$ contribution shows no sensitivity to dynamical off-shellness. However this is, at the kinematics of MiniBoone, a relatively small contribution for the neutrino case. The $T$ and $T'$ contributions are the dominant components of the cross section, and they exhibit a similar effect of off-shellness: off-shell effects tend to increase (in absolute magnitude) both $T$ and $T'$ contributions. Further the $T$ contribution is the same for neutrino and antineutrino, while the $T'$ changes sign. As a consequence, while for neutrinos off-shell effects in the $T$ and $T'$ contributions (which add) are reinforced and a net visible dependence of the off-shellness is seen in the total cross section, for antineutrinos such effects are nearly perfectly canceled in the cross section at MiniBooNE kinematics, since $T$ and $T'$ contributions tend to cancel.

\begin{figure}[h]\centering
\includegraphics[width=0.6\columnwidth]{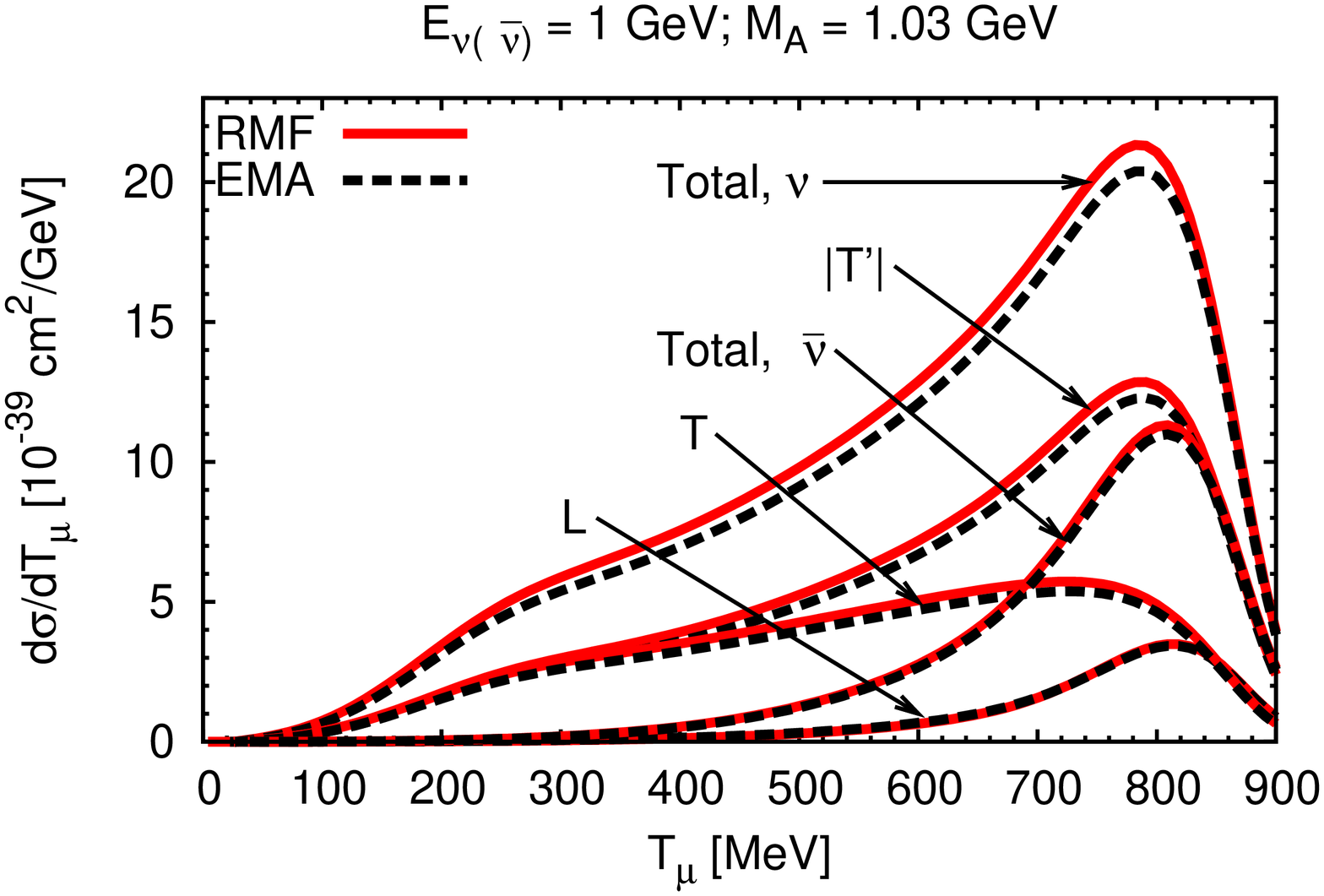}
\caption{(Color online) The differential cross section per target nucleon for the (anti)neutrino
CCQE process on $^{12}$C as a function of muon kinetic energy $T_\mu$, as well as transverse ($T$),
longitudinal ($L$), and transverse-transverse ($|T'|$) contributions using the EMA approach
(dashed, black online) and the RMF model (solid, red online).\label{fig00}}
\end{figure}

\begin{figure}[h]\centering
\includegraphics[width=0.6\columnwidth]{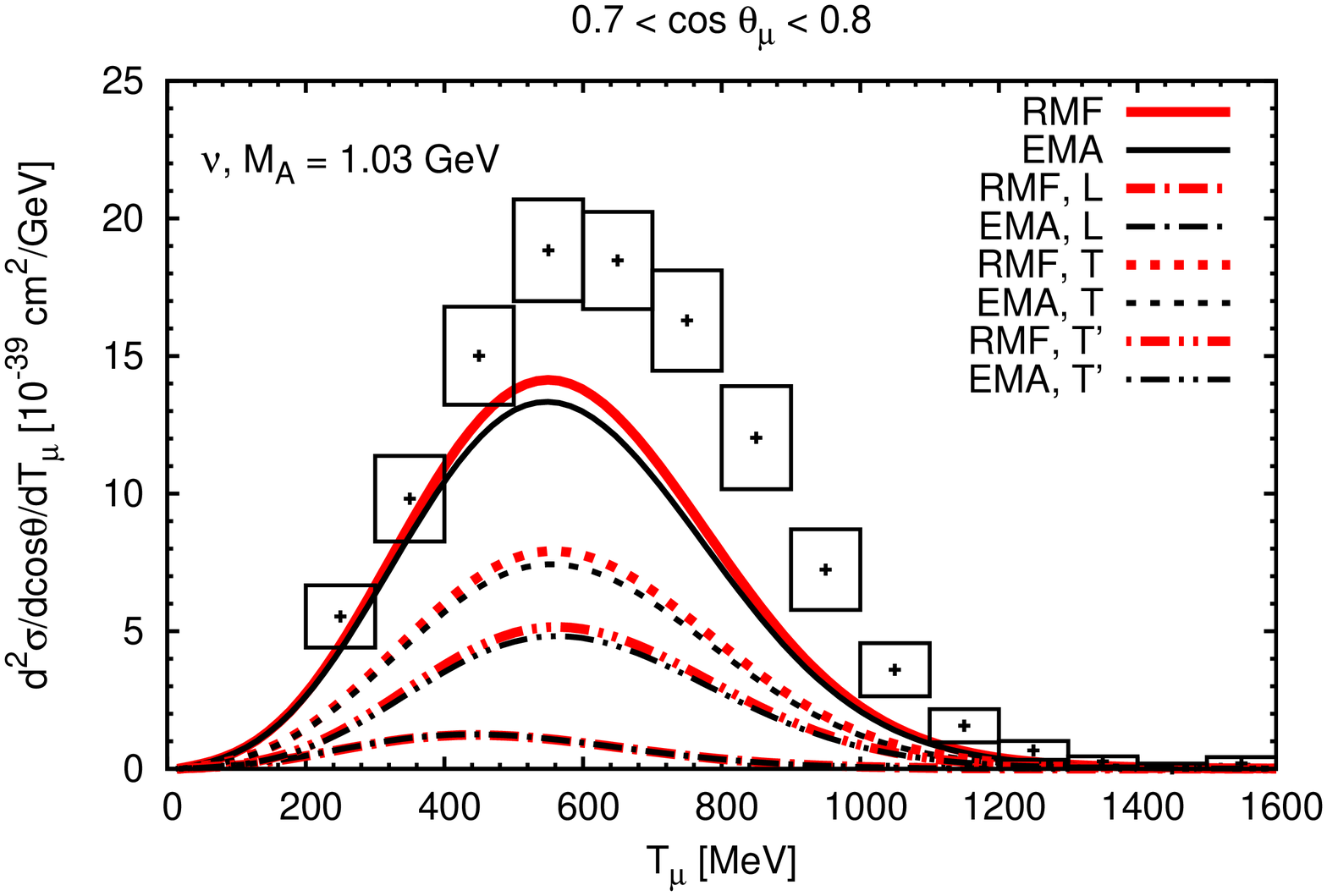}\\[5pt]
\includegraphics[width=0.6\columnwidth]{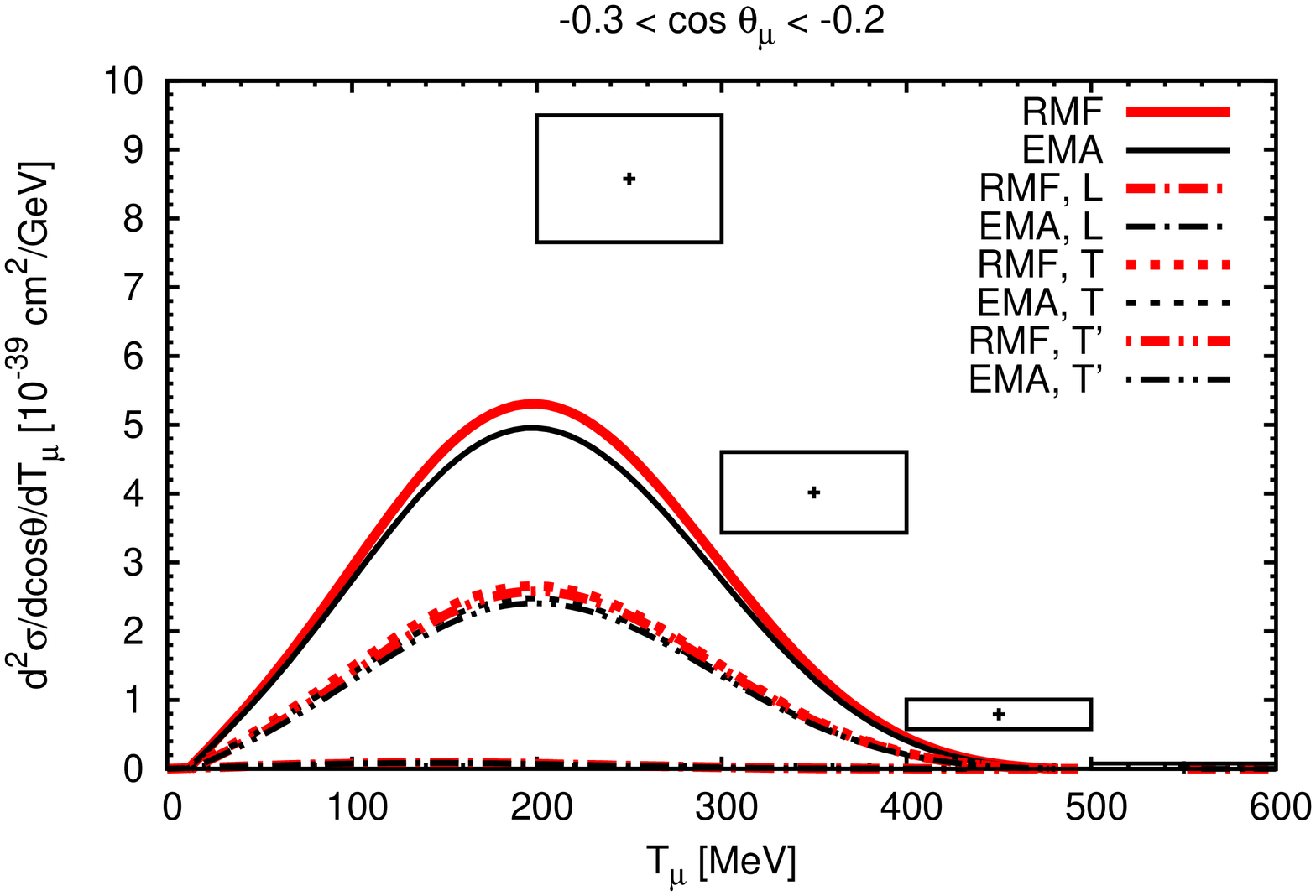}
\caption{(Color online) Flux-integrated double-differential cross section per target nucleon for the $\nu_\mu$ CCQE process on $^{12}$C displayed versus the $\mu^-$ kinetic energy $T_\mu$ for two bins of $\cos\theta_\mu$ (forward angles -- top panel and backward angles -- bottom angles) obtained within the RMF model (solid thick line), EMA approach (solid thin line) and contribution of longitudinal ($\sigma_L$, dash-dotted line) and transverse ($\sigma_T$ -- dotted line and $\sigma_{T'}$ -- dash-dot-dot line) components within RMF and EMA models. The data are from \cite{AguilarArevalo:2010zc}.}\label{fig01}
\end{figure}

\begin{figure}[h]\centering
\includegraphics[width=0.6\columnwidth]{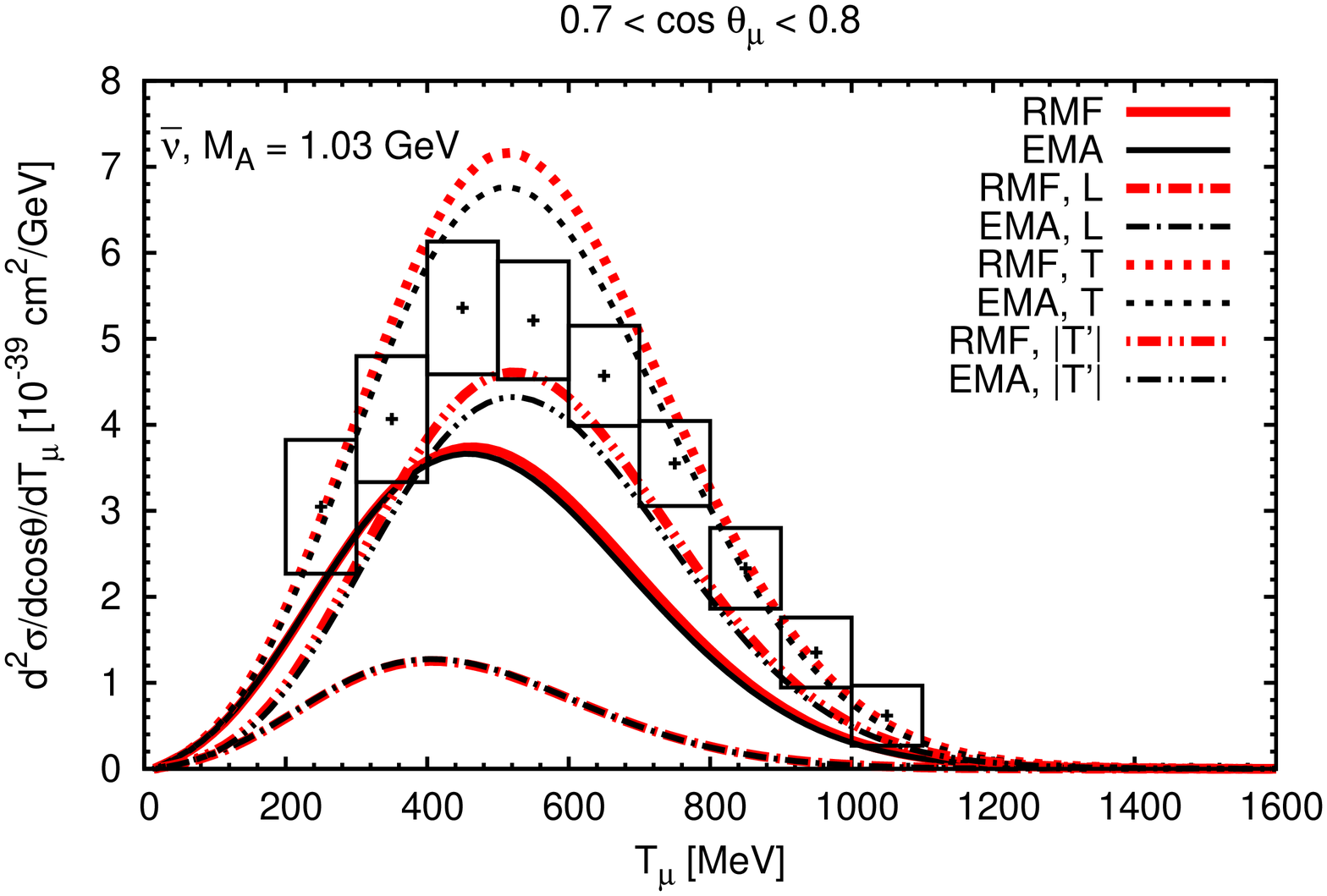}\\[5pt]
\includegraphics[width=0.6\columnwidth]{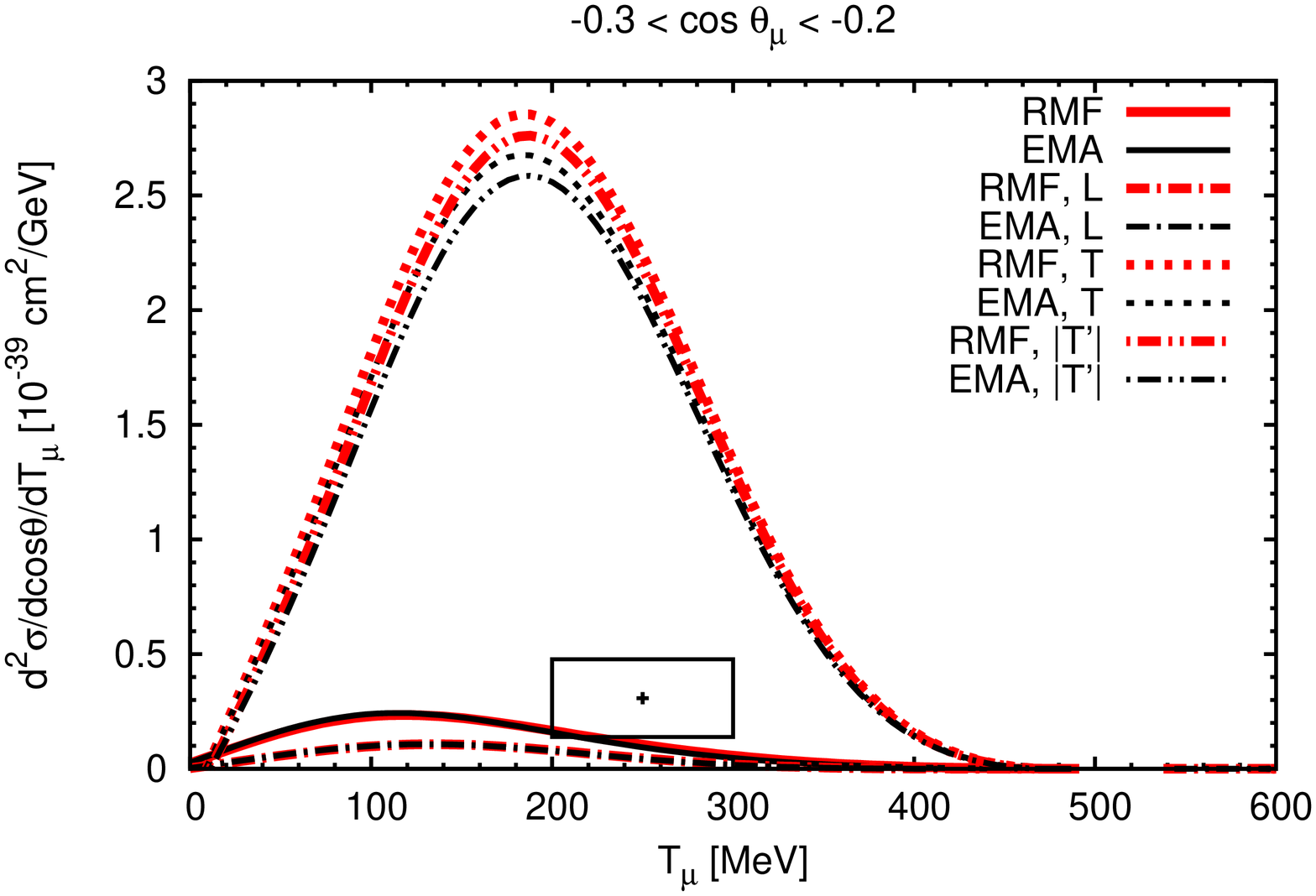}
\caption{(Color online) As for Fig.~\ref{fig01}, but for $\overline\nu_\mu$ scattering versus $\mu^+$ kinetic energy $T_\mu$: RMF model (solid thick line), EMA approach (solid thin line) and contribution of longitudinal ($\sigma_L$, dash-dotted line) and transverse ($\sigma_T$ -- dotted line and $\sigma_{T'}$ -- dash-dot-dot line) components within RMF and EMA models. The data are from \cite{CCQE-anti-exp}.}\label{fig02}
\end{figure}

\begin{figure*}[t]\centering
\includegraphics[width=.33\columnwidth]{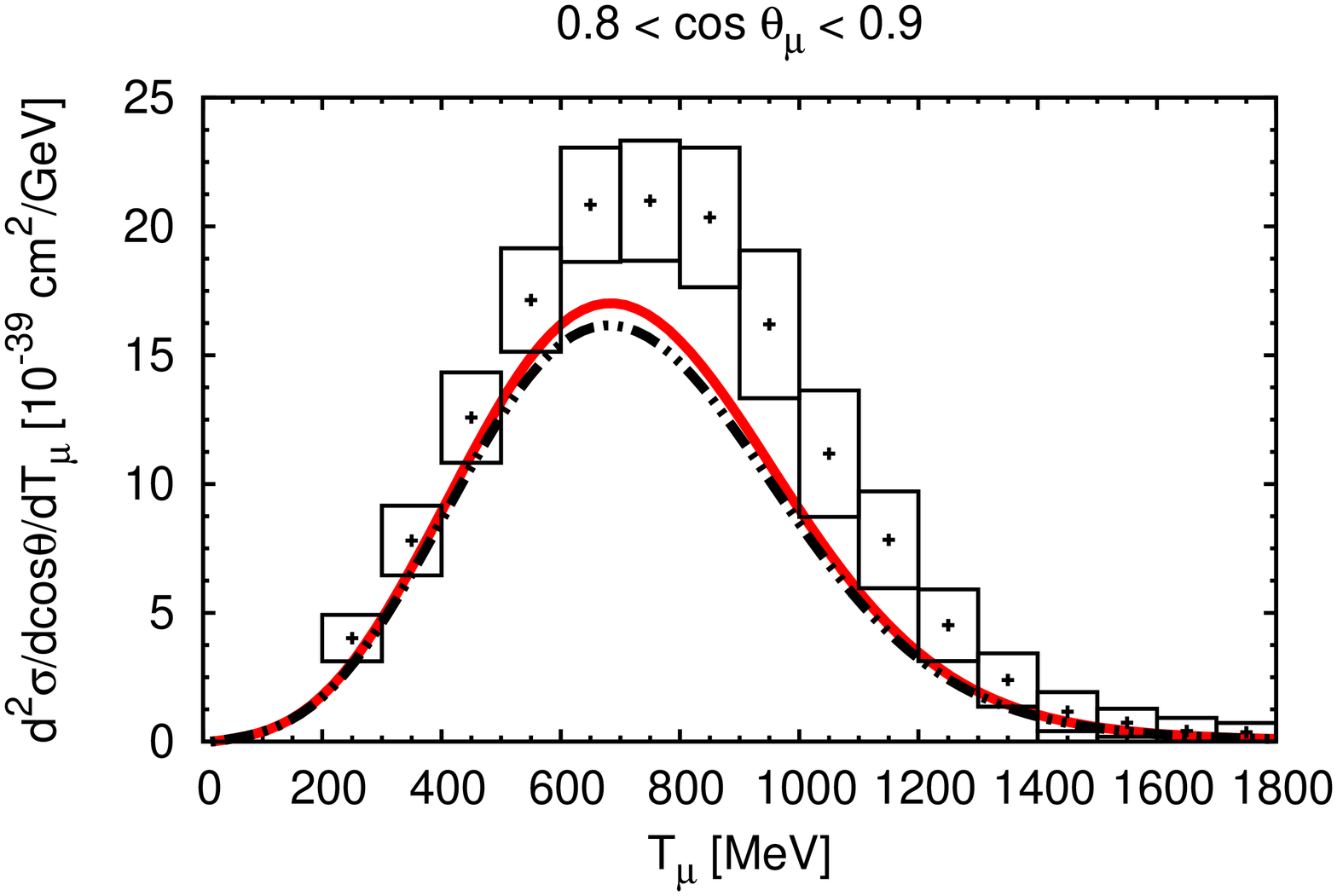}\includegraphics[width=.33\columnwidth]{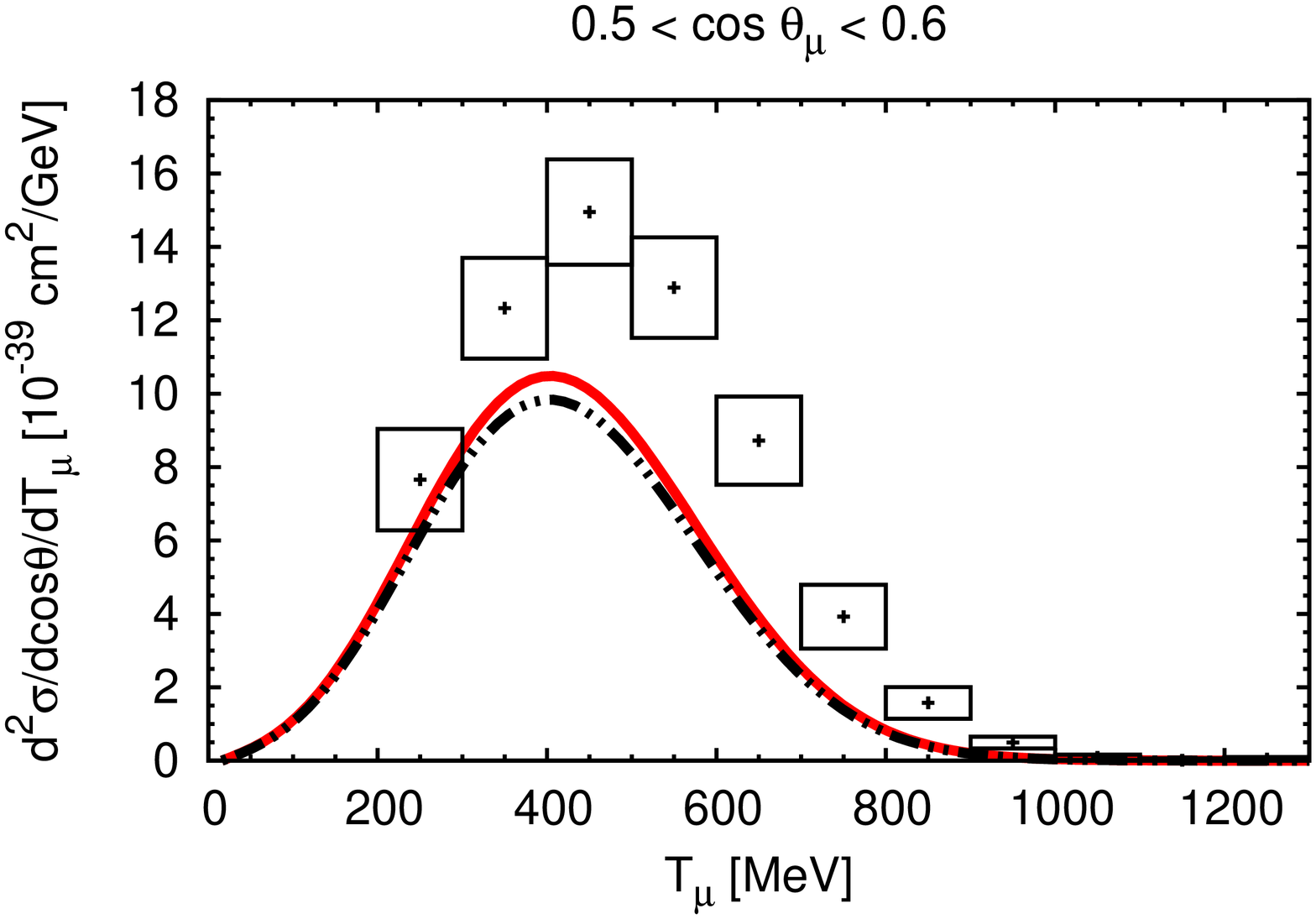}\includegraphics[width=.33\columnwidth]{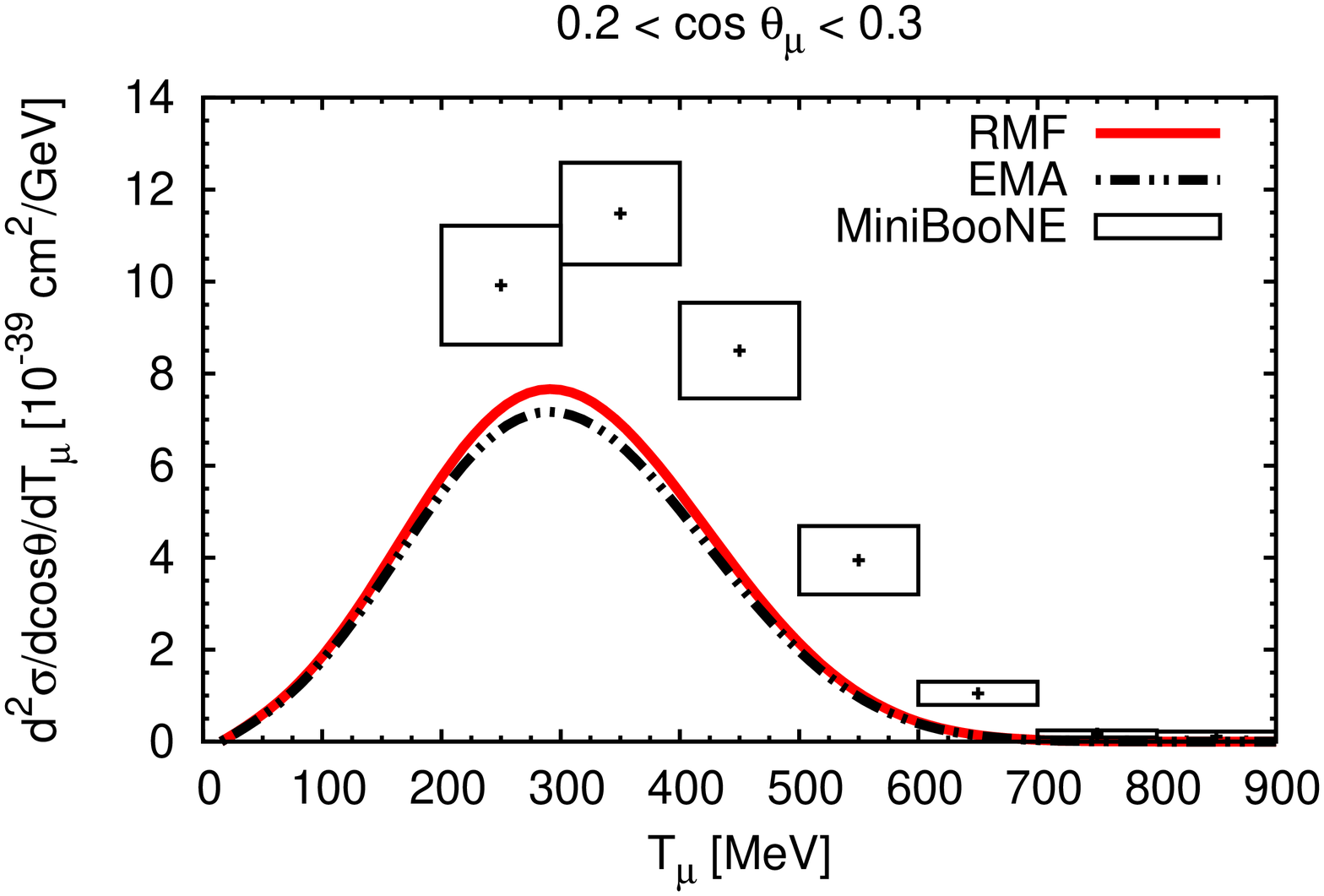}\\[5pt]
\includegraphics[width=.33\columnwidth]{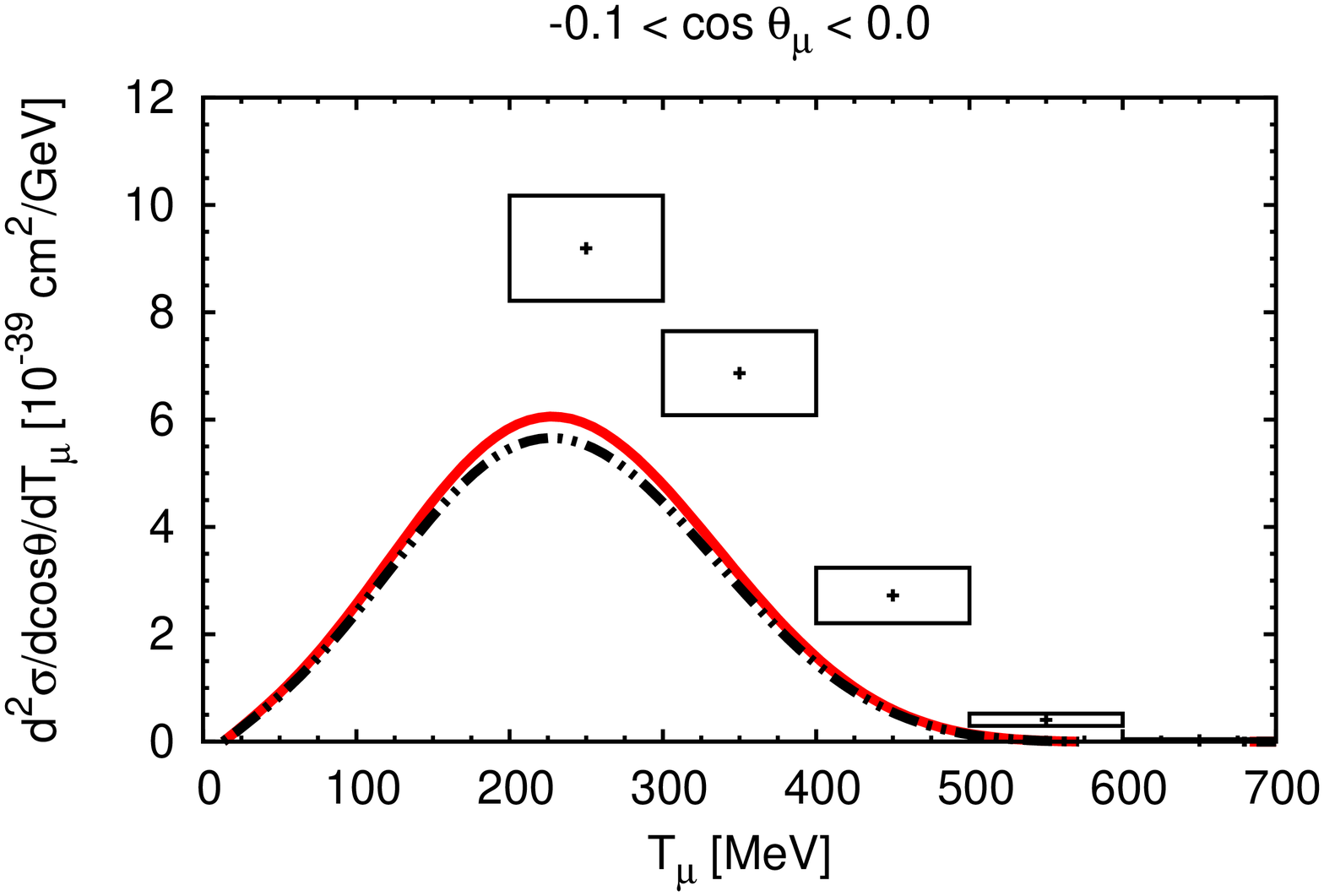}\includegraphics[width=.33\columnwidth]{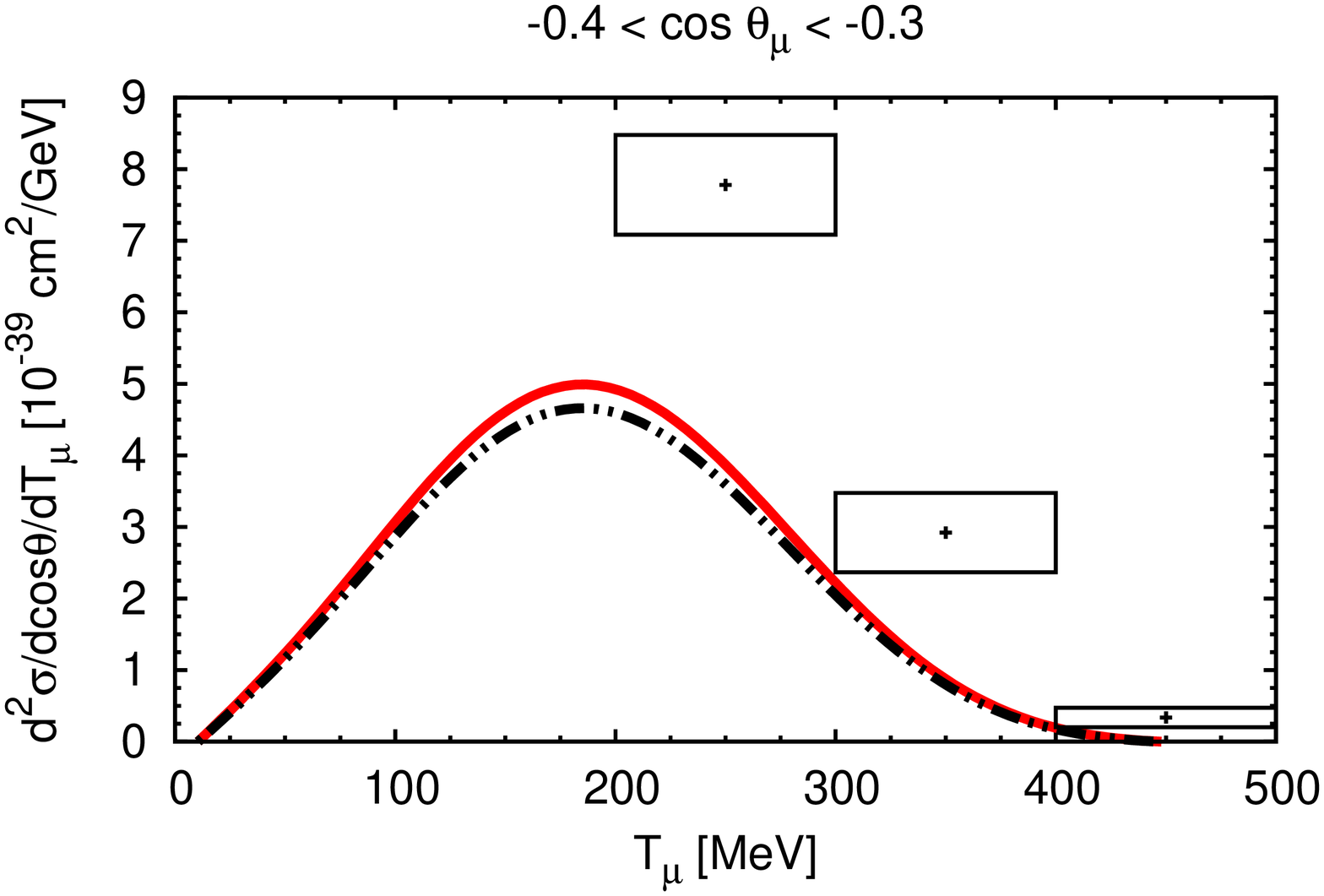}\includegraphics[width=.33\columnwidth]{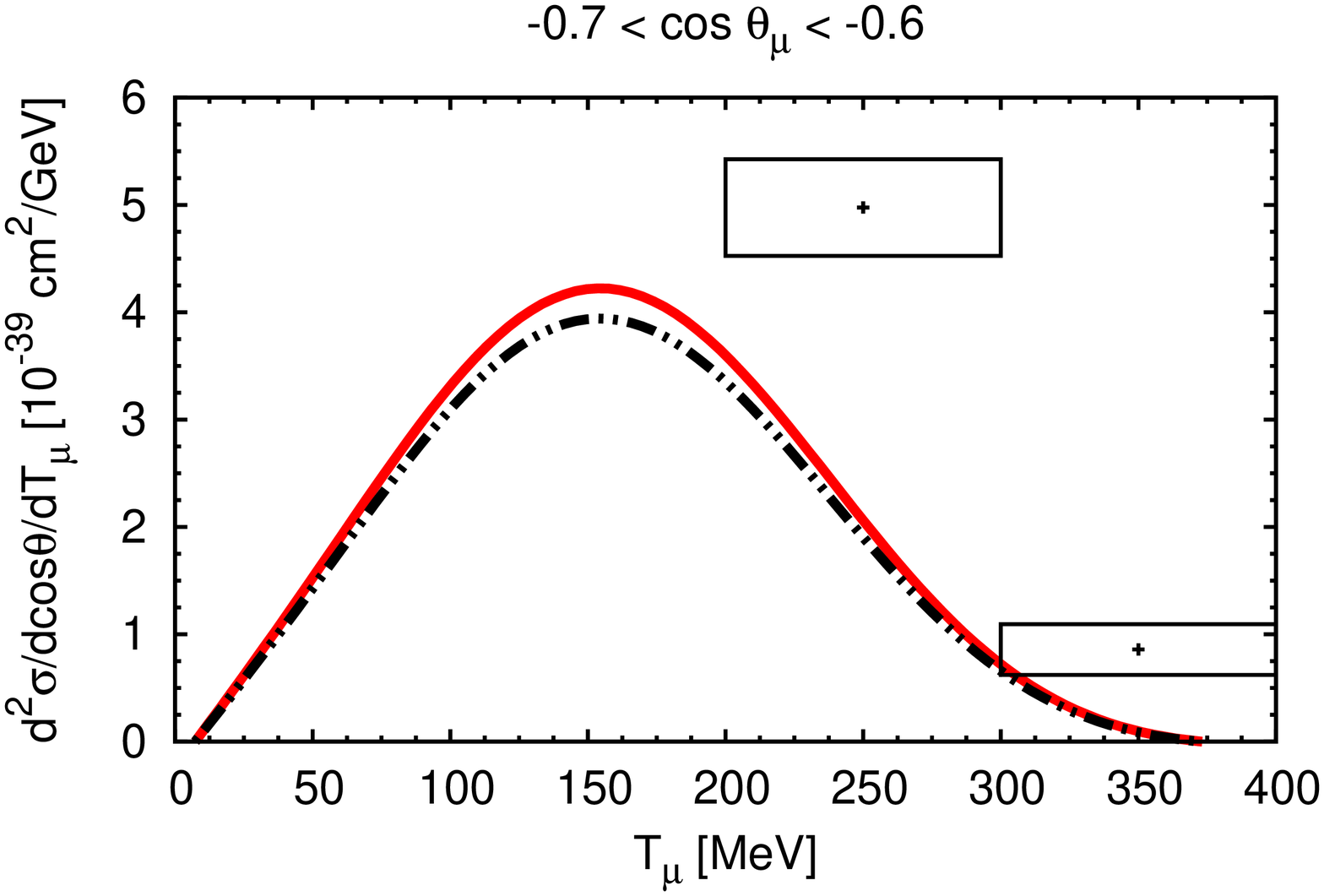}
\caption{(Color online) Flux-integrated double-differential cross section per target nucleon for the $\nu_\mu$ CCQE process on $^{12}$C displayed versus the $\mu^-$ kinetic energy $T_\mu$ for various bins of $\cos\theta_\mu$ obtained within the RMF model and EMA approach for $M_A=1.03$. The data are from \cite{AguilarArevalo:2010zc}.}\label{fig03}
\end{figure*}

\begin{figure*}[t]\centering
\includegraphics[width=.33\columnwidth]{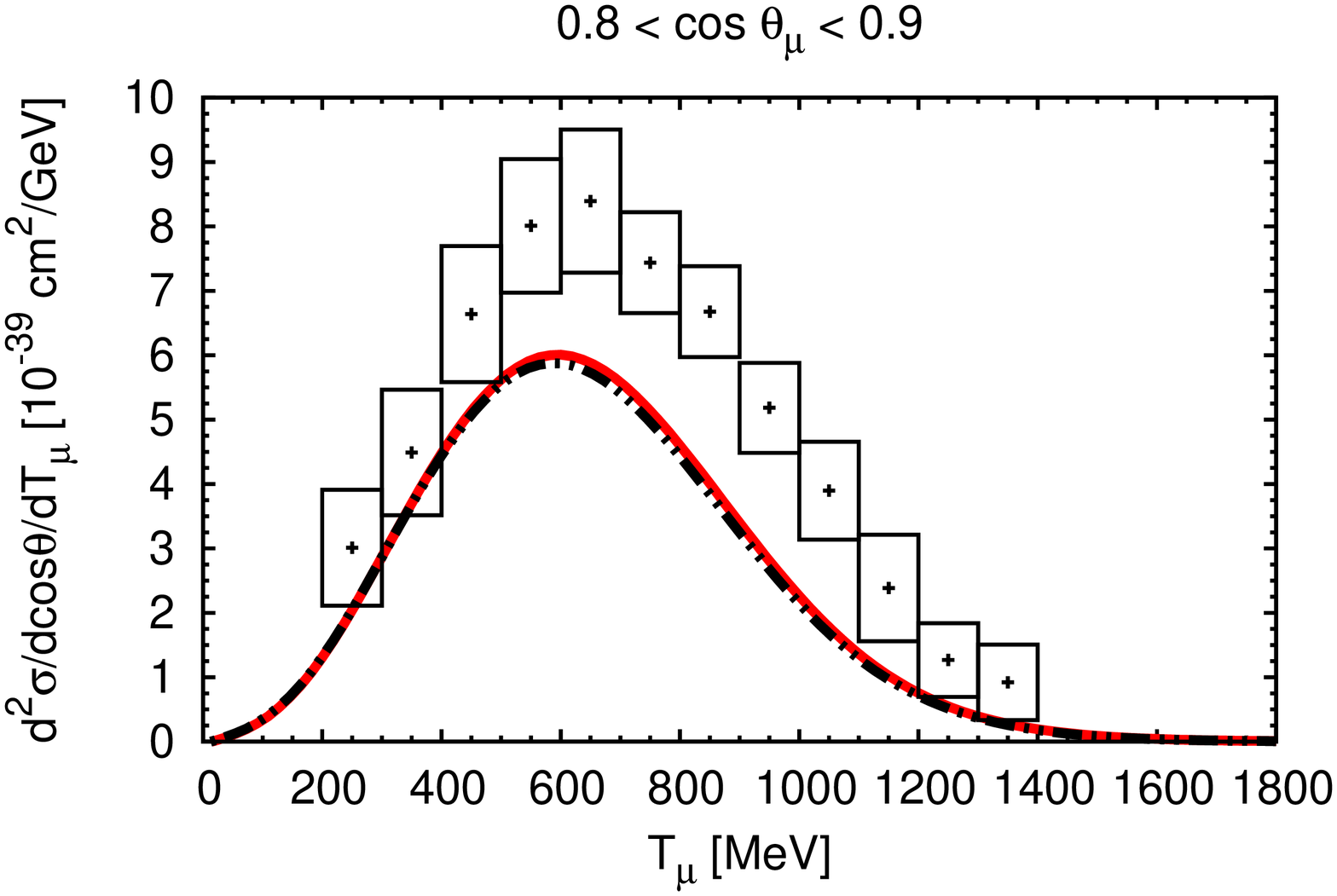}\includegraphics[width=.33\columnwidth]{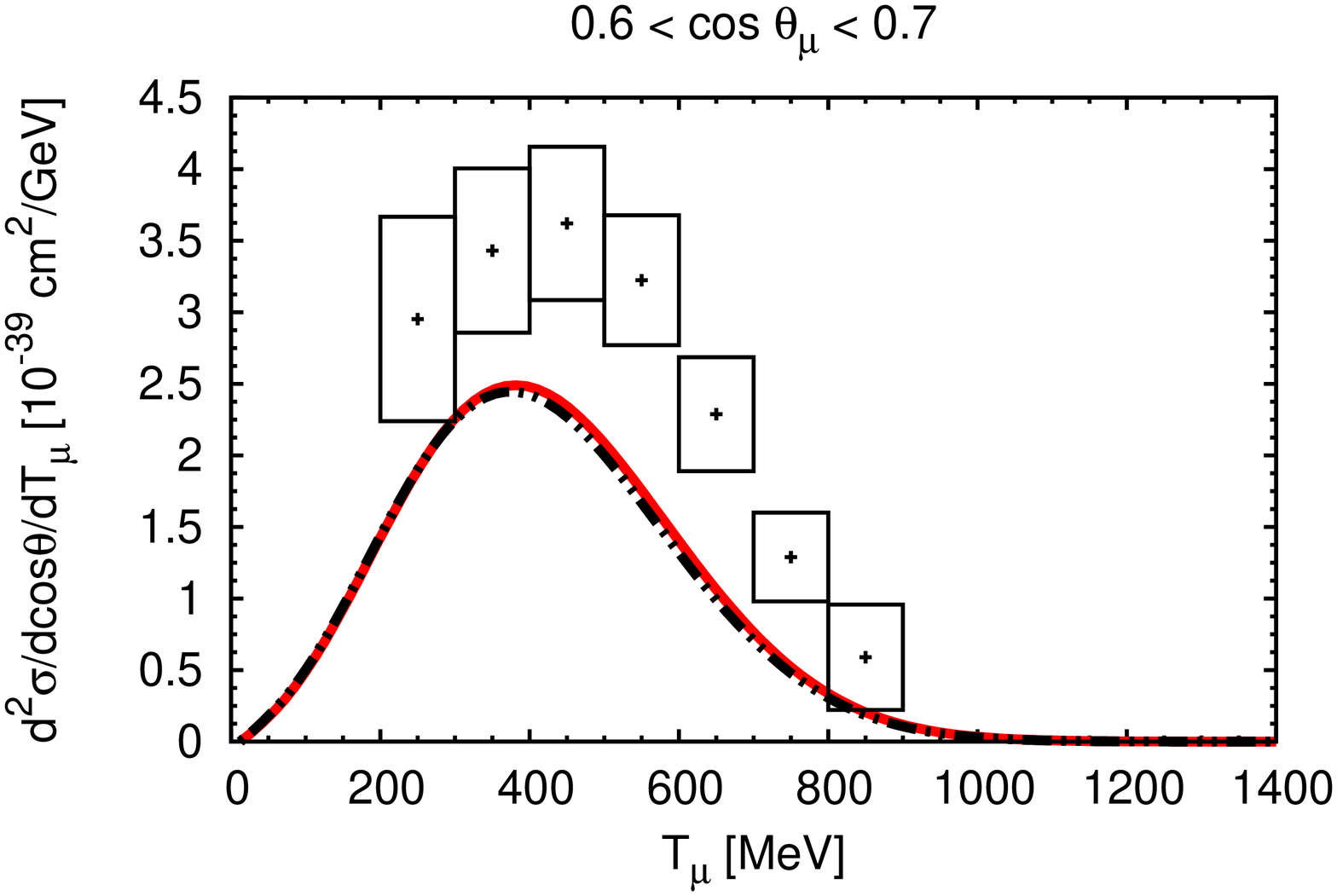}\includegraphics[width=.33\columnwidth]{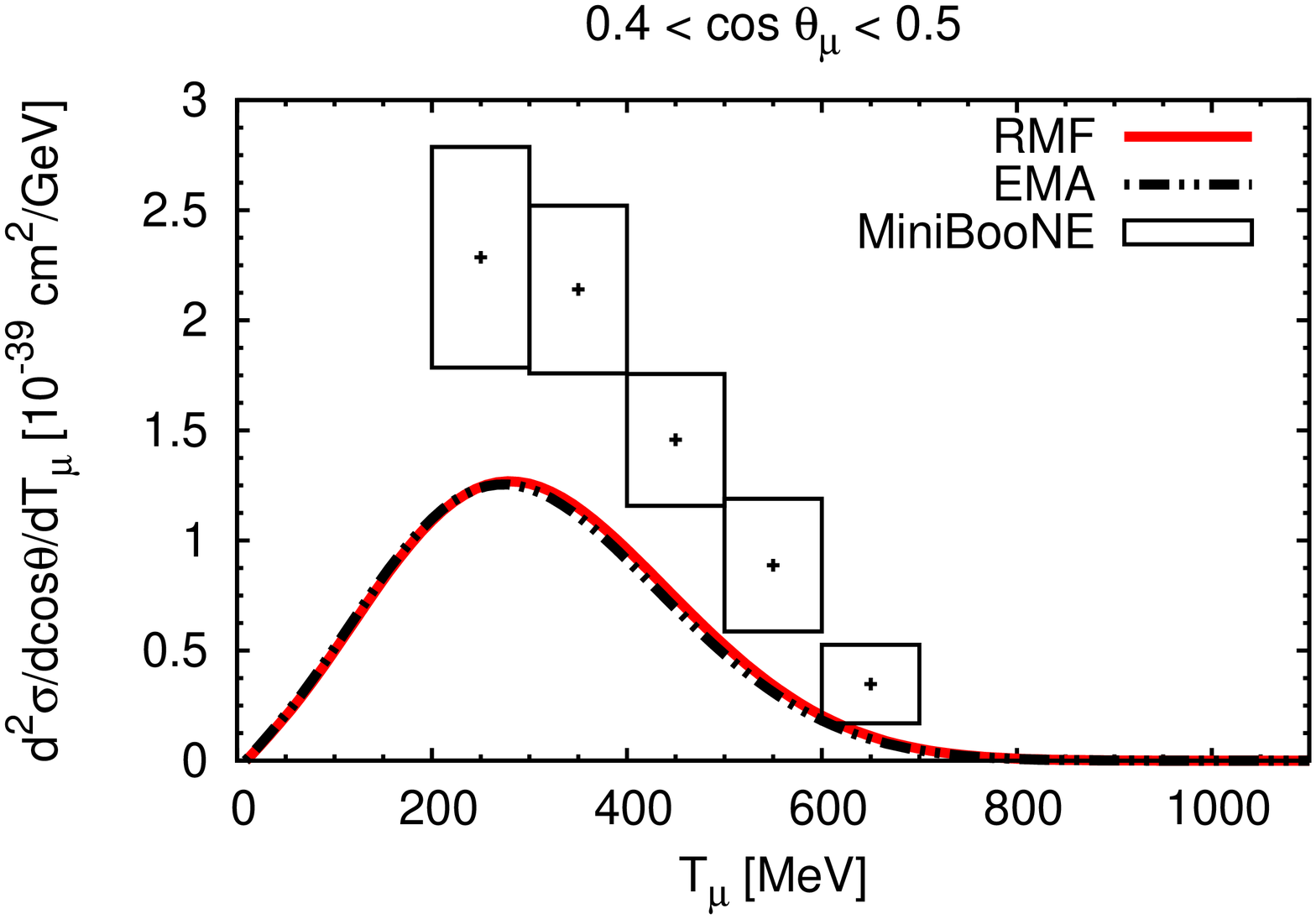}\\[5pt]
\includegraphics[width=.33\columnwidth]{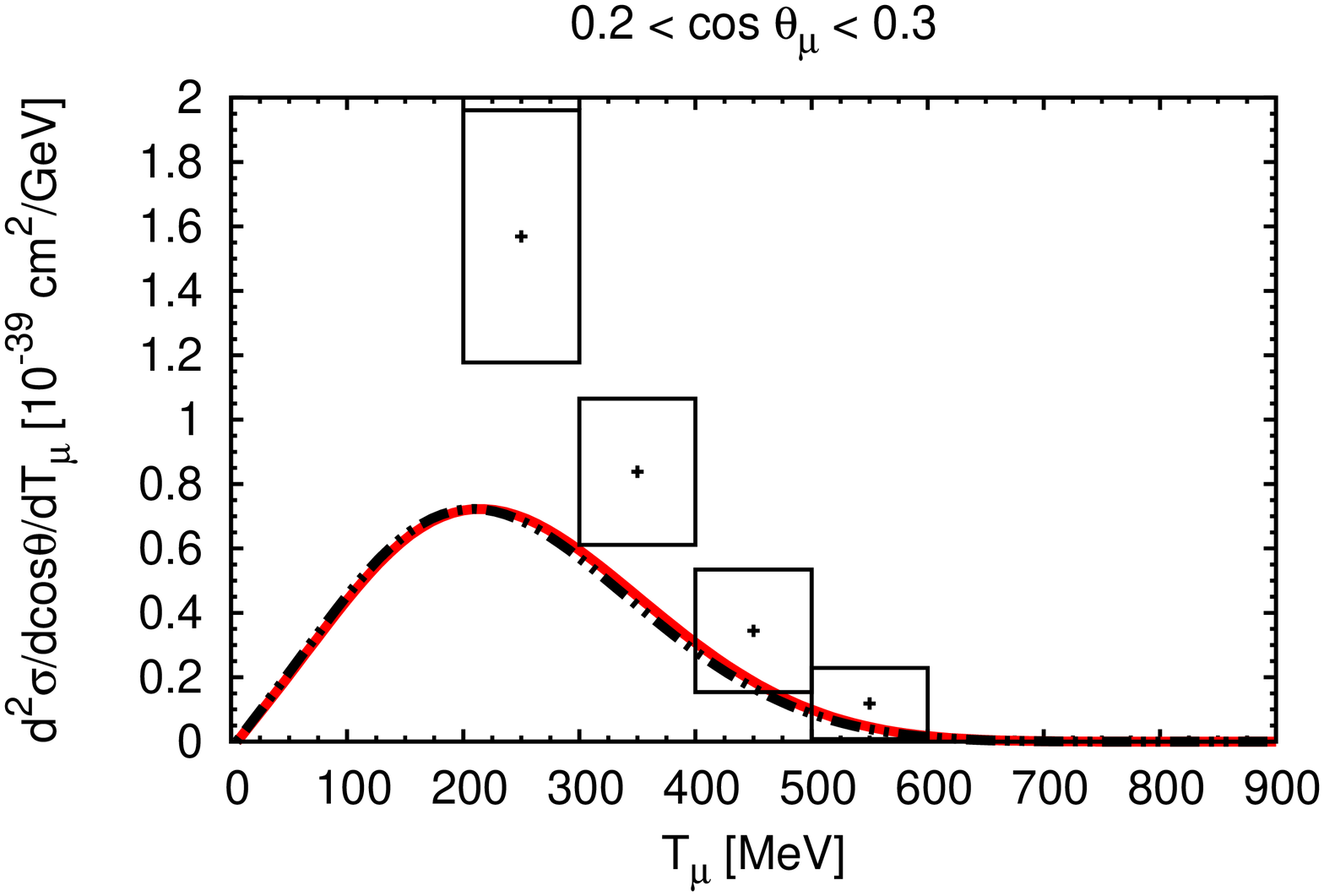}\includegraphics[width=.33\columnwidth]{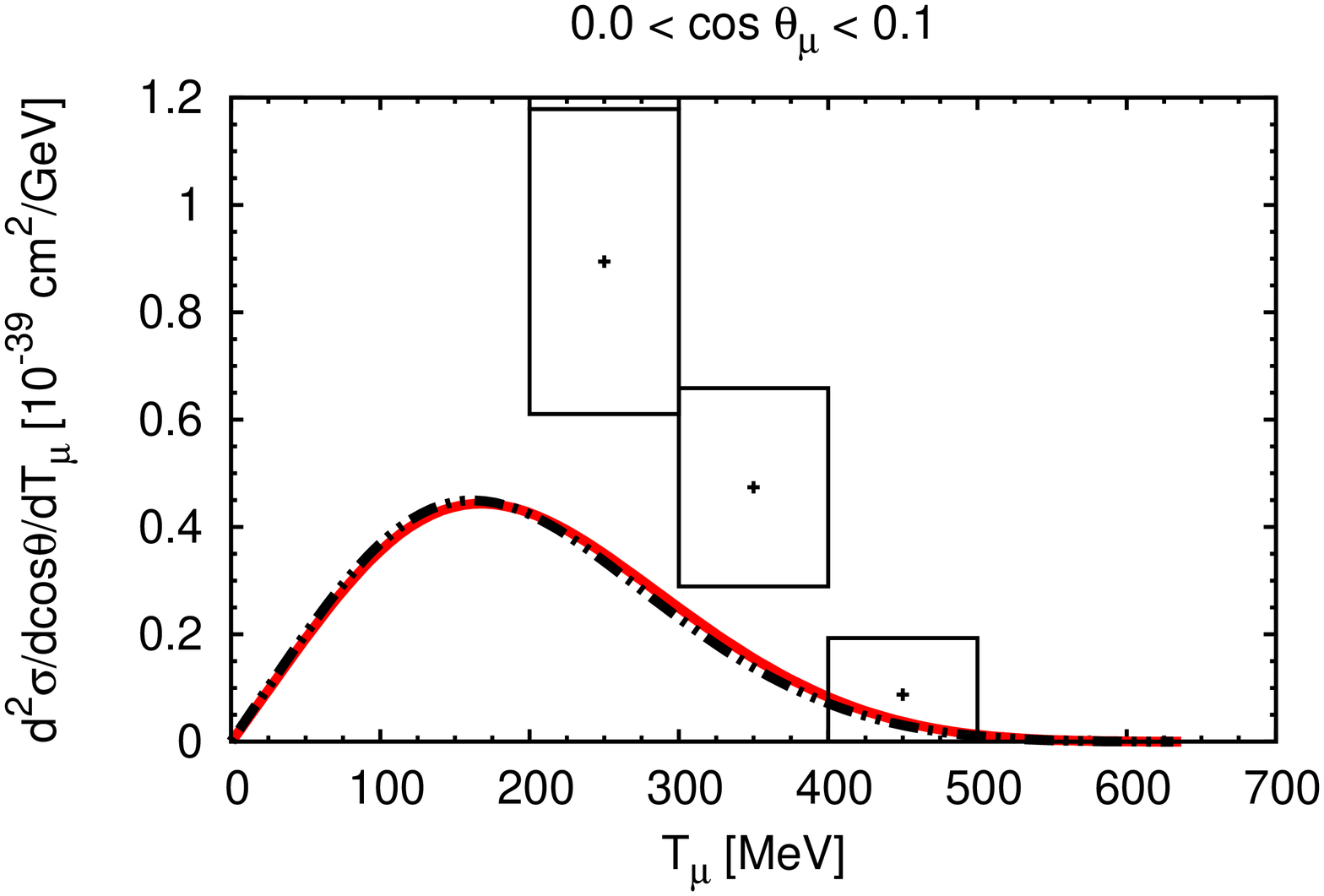}\includegraphics[width=.33\columnwidth]{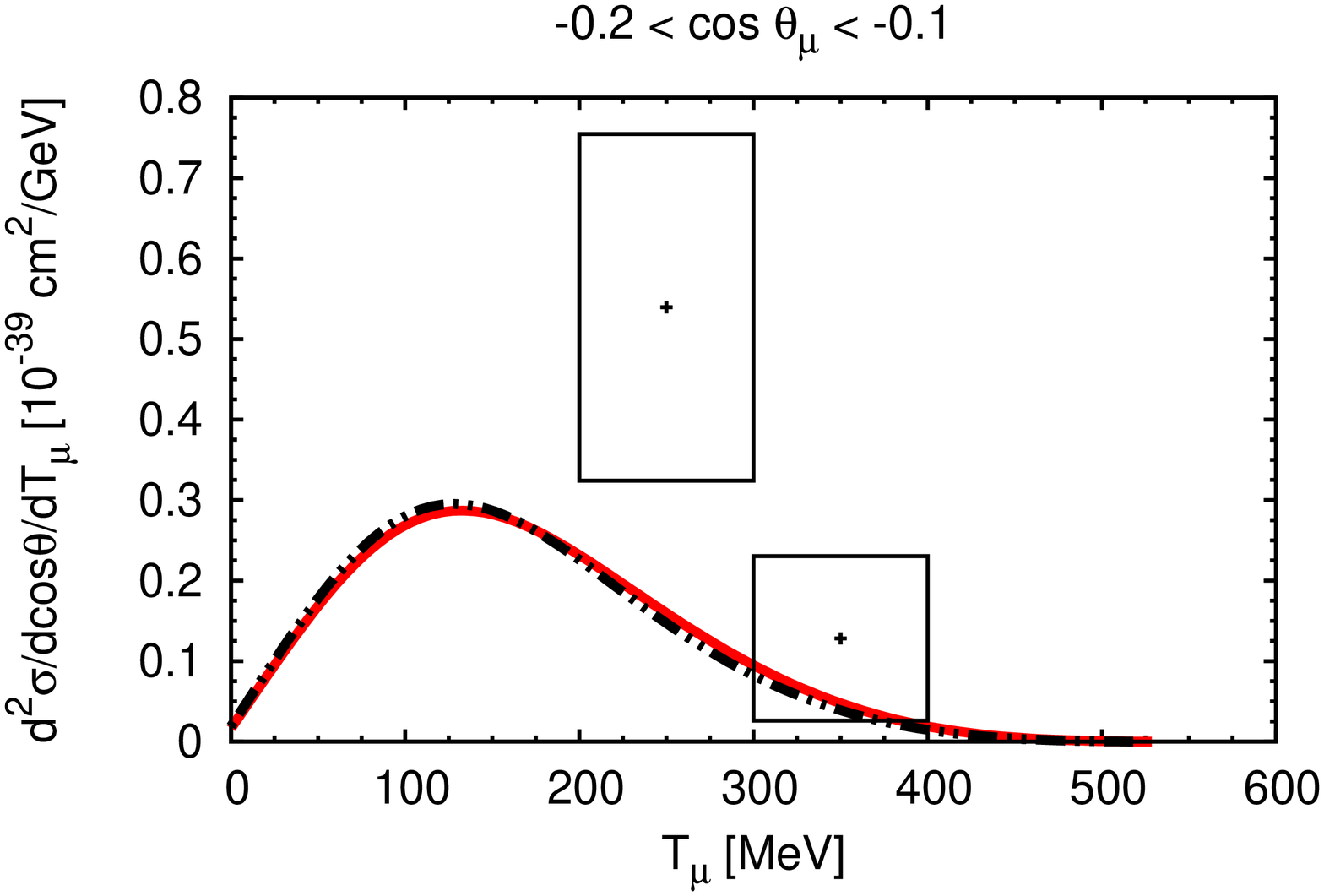}
\caption{(Color online) As for Fig.~\ref{fig03}, but for $\overline\nu_\mu$ scattering versus $\mu^+$ kinetic energy $T_\mu$. The data are from \cite{CCQE-anti-exp}.}\label{fig04}
\end{figure*}

\begin{figure*}[t]\centering
\includegraphics[height=45mm]{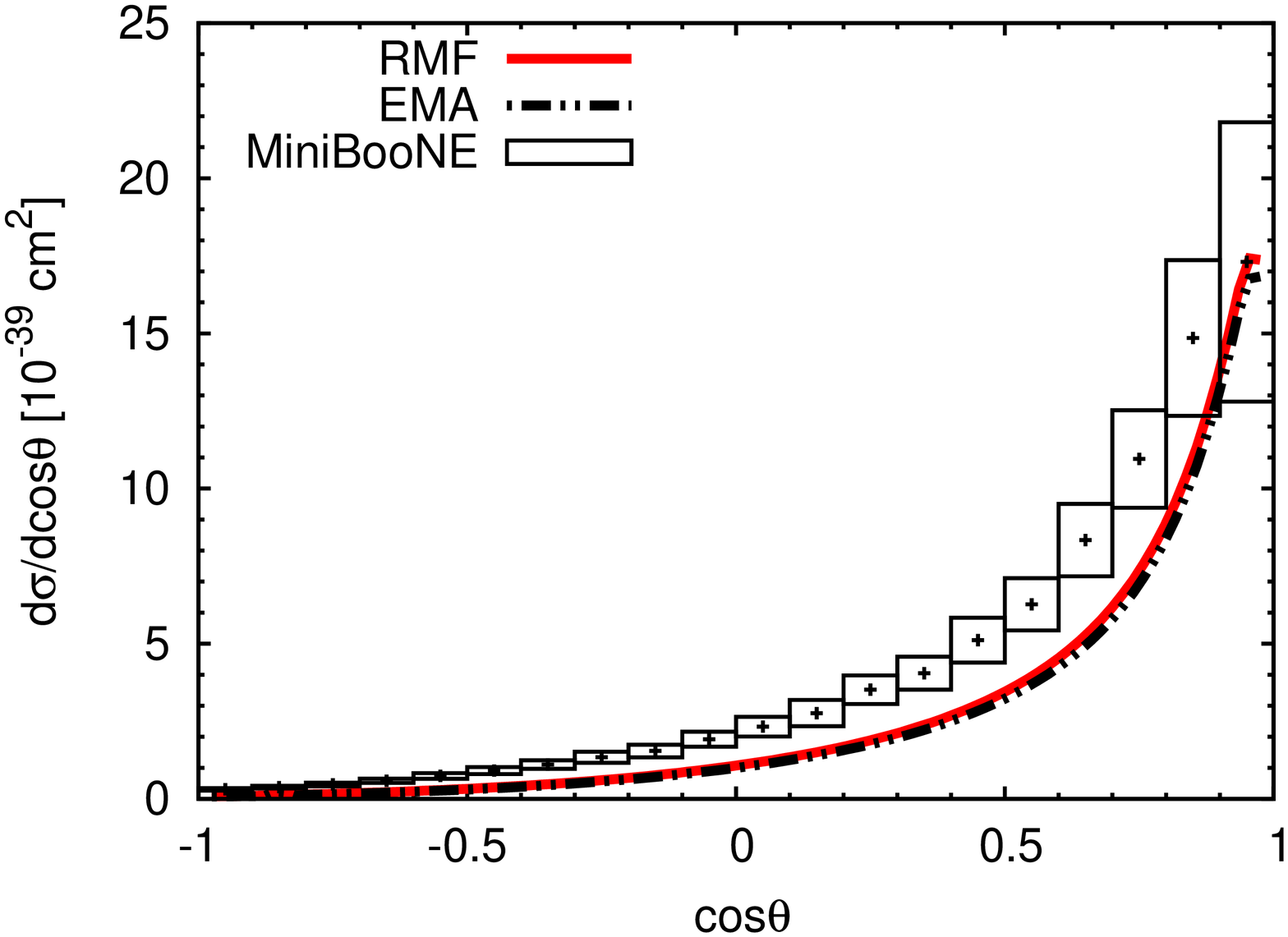}\hspace*{5pt}\includegraphics[height=45mm]{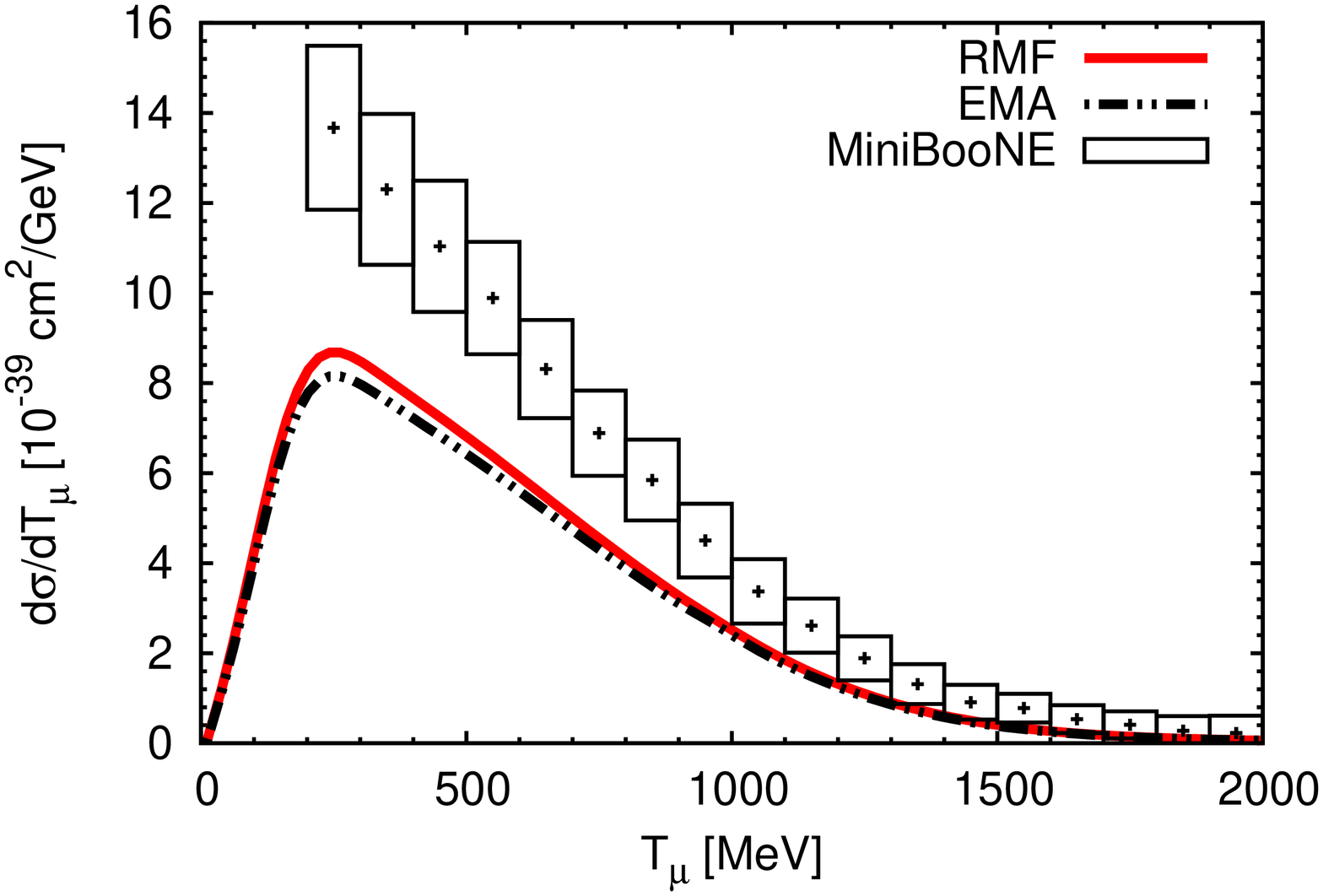}\\[5pt]
\includegraphics[height=45mm]{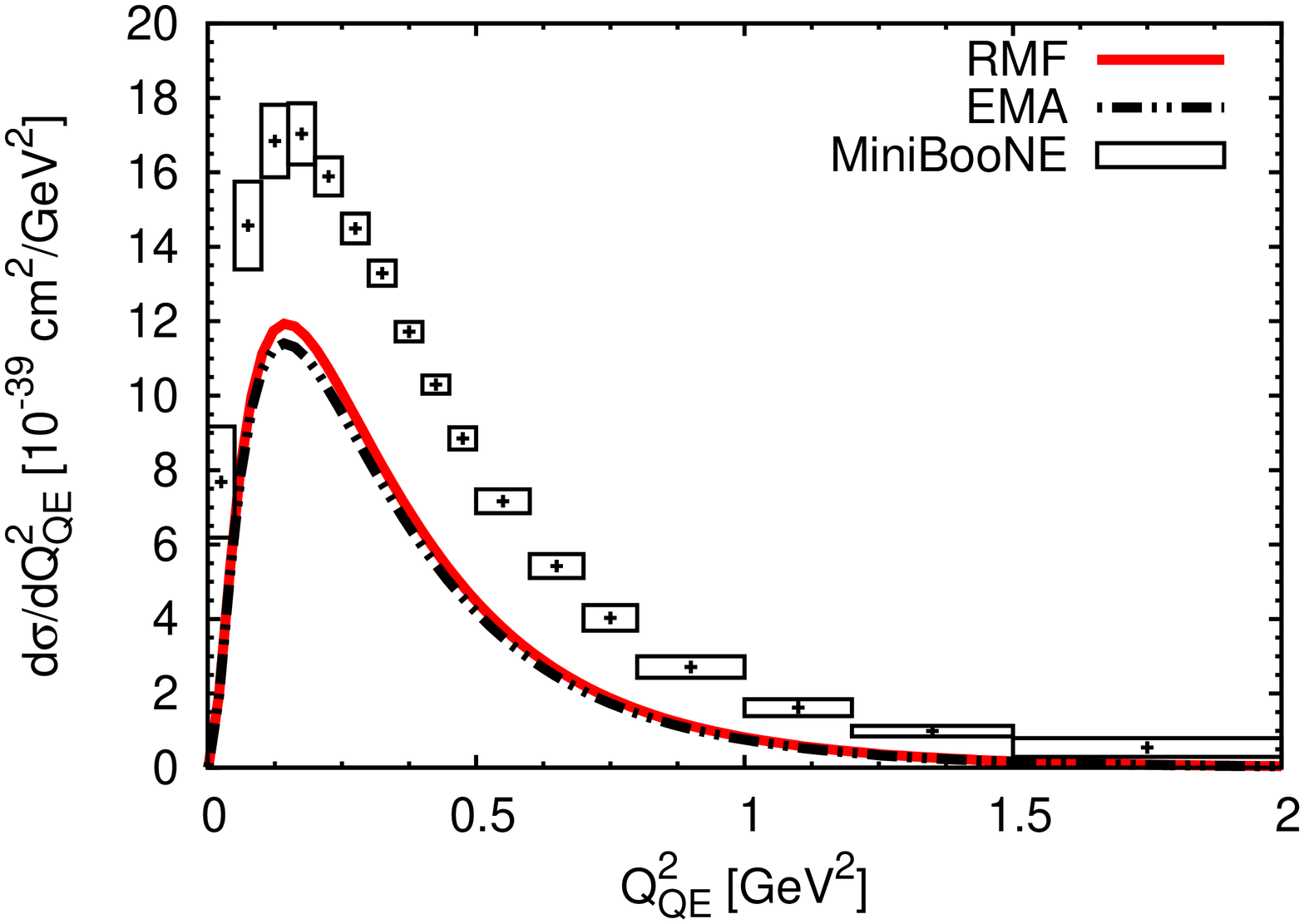}\hspace*{5pt}\includegraphics[height=45mm]{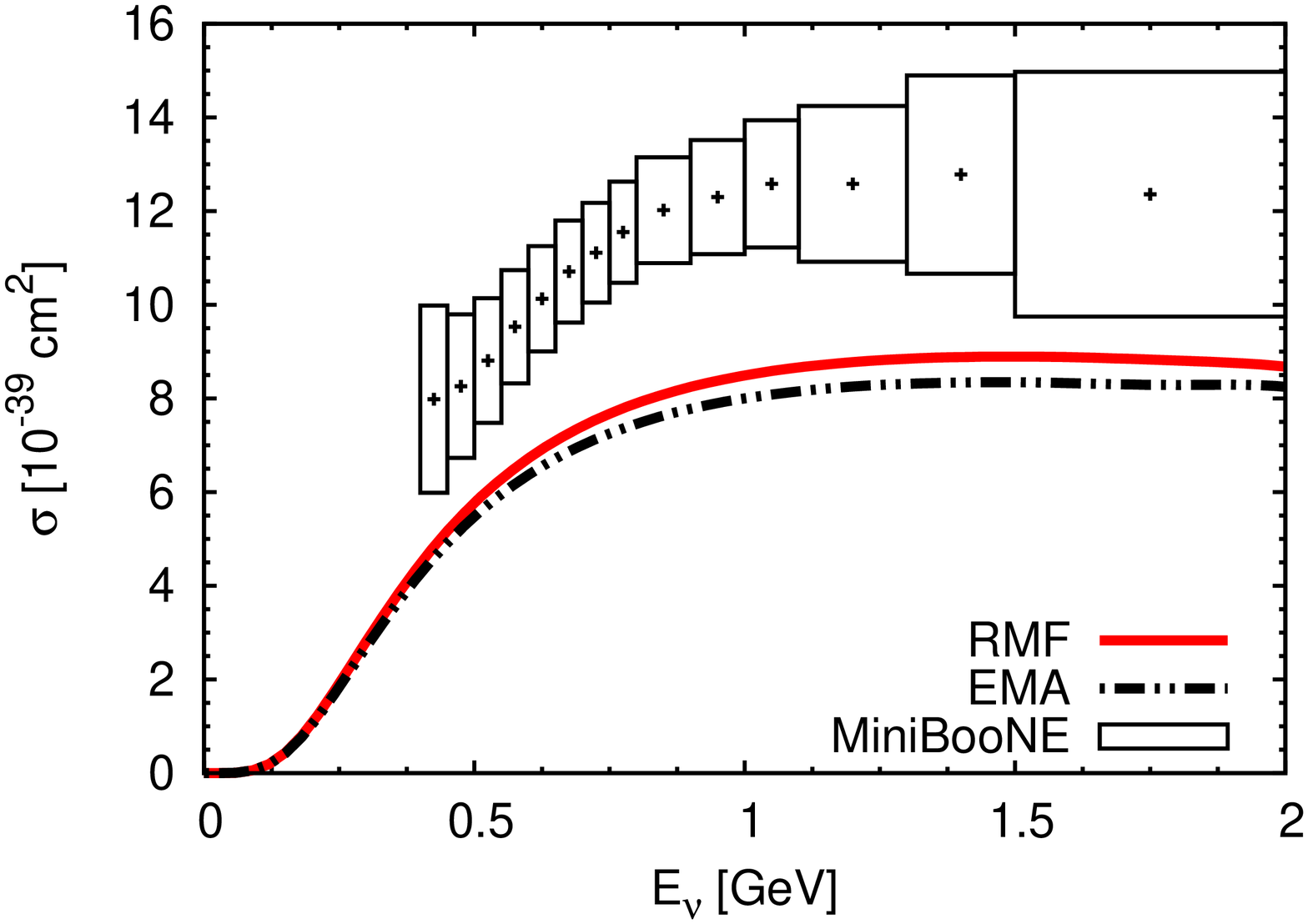}
\caption{(Color online) MiniBooNE flux-averaged CCQE $\nu_\mu$-$^{12}$C differential cross section per neutron as a function of the muon scattering angle (top-left panel), of the muon kinetic energy (top-right panel), of the four momentum transfer $Q^2$  (bottom-left panel) and total CCQE $\nu_\mu$-$^{12}$C cross section per neutron as a function of neutrino energy (bottom-right panel). Results are obtained within RMF model and EMA approach. The data are from \cite{AguilarArevalo:2010zc}.}\label{fig05}
\end{figure*}

\begin{figure*}[t]\centering
\includegraphics[height=45mm]{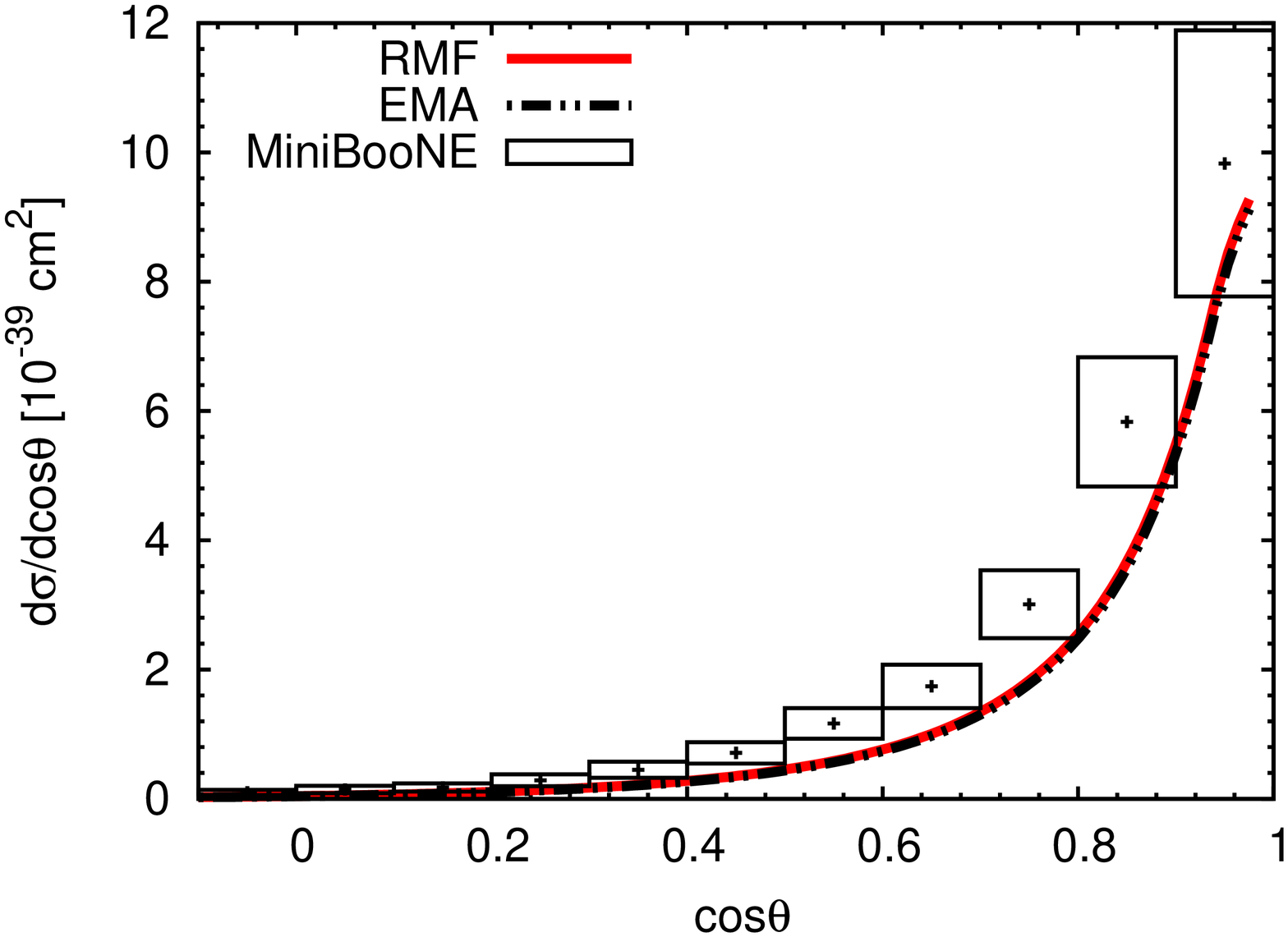}\hspace*{5pt}\includegraphics[height=45mm]{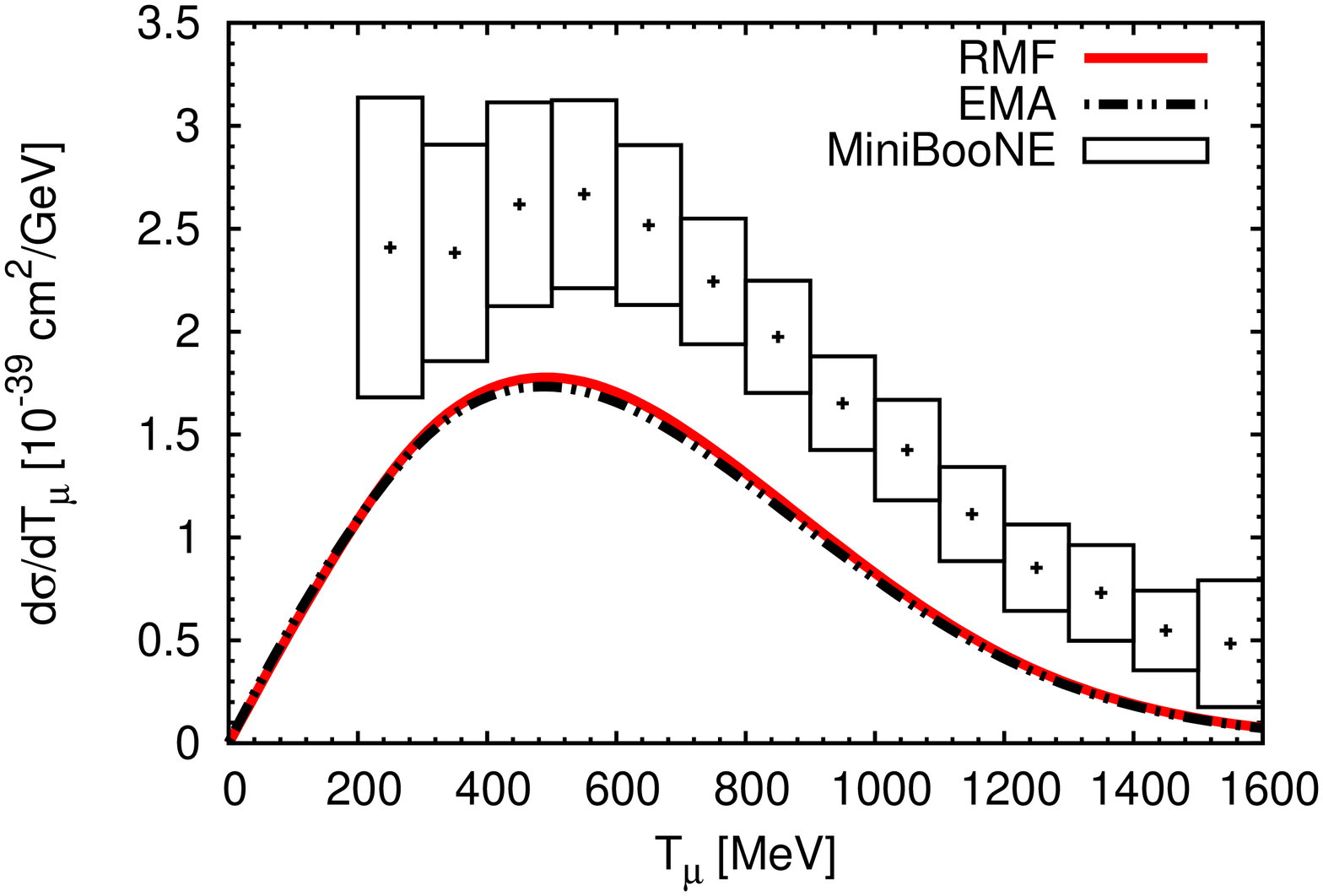}\\[5pt]
\includegraphics[height=45mm]{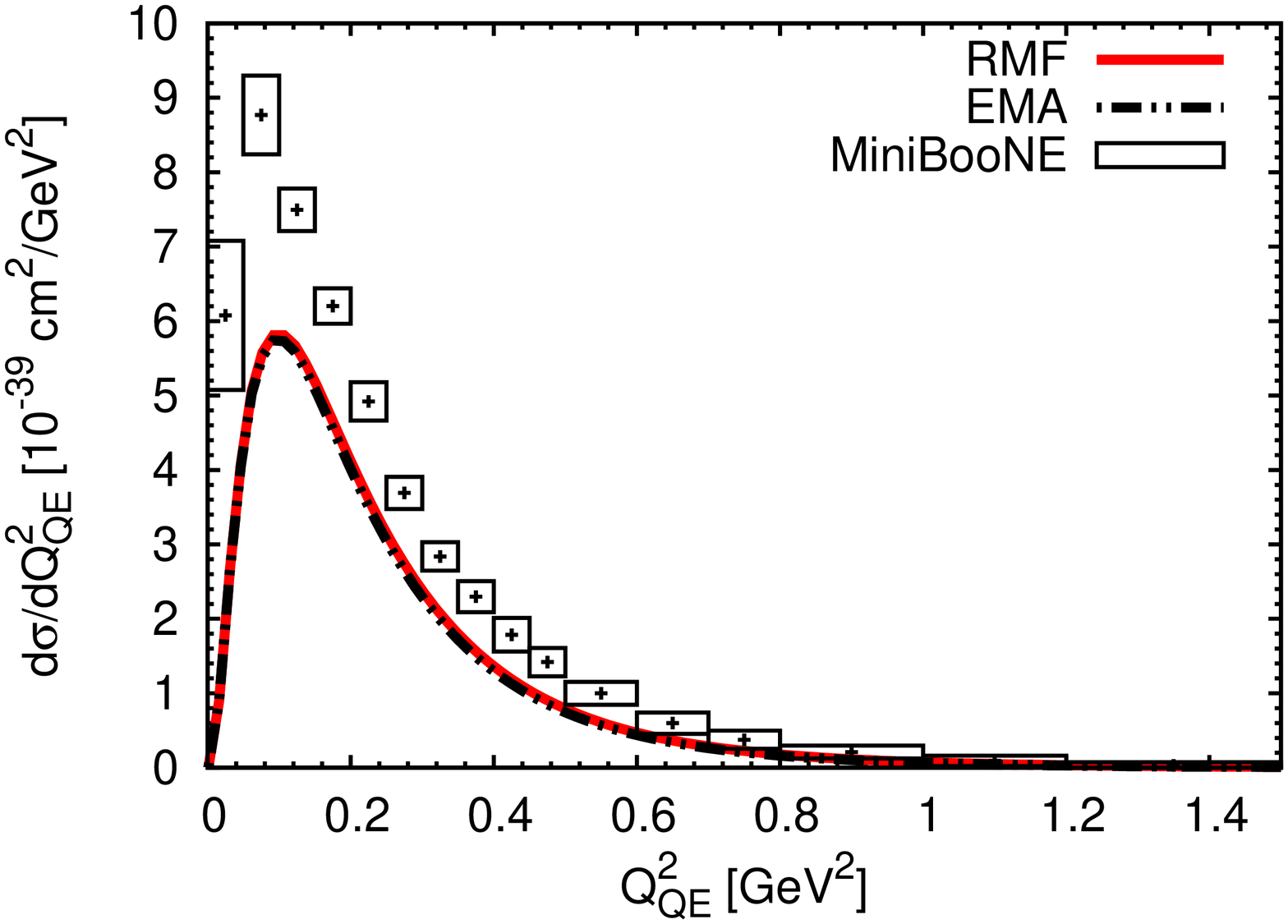}\hspace*{5pt}\includegraphics[height=45mm]{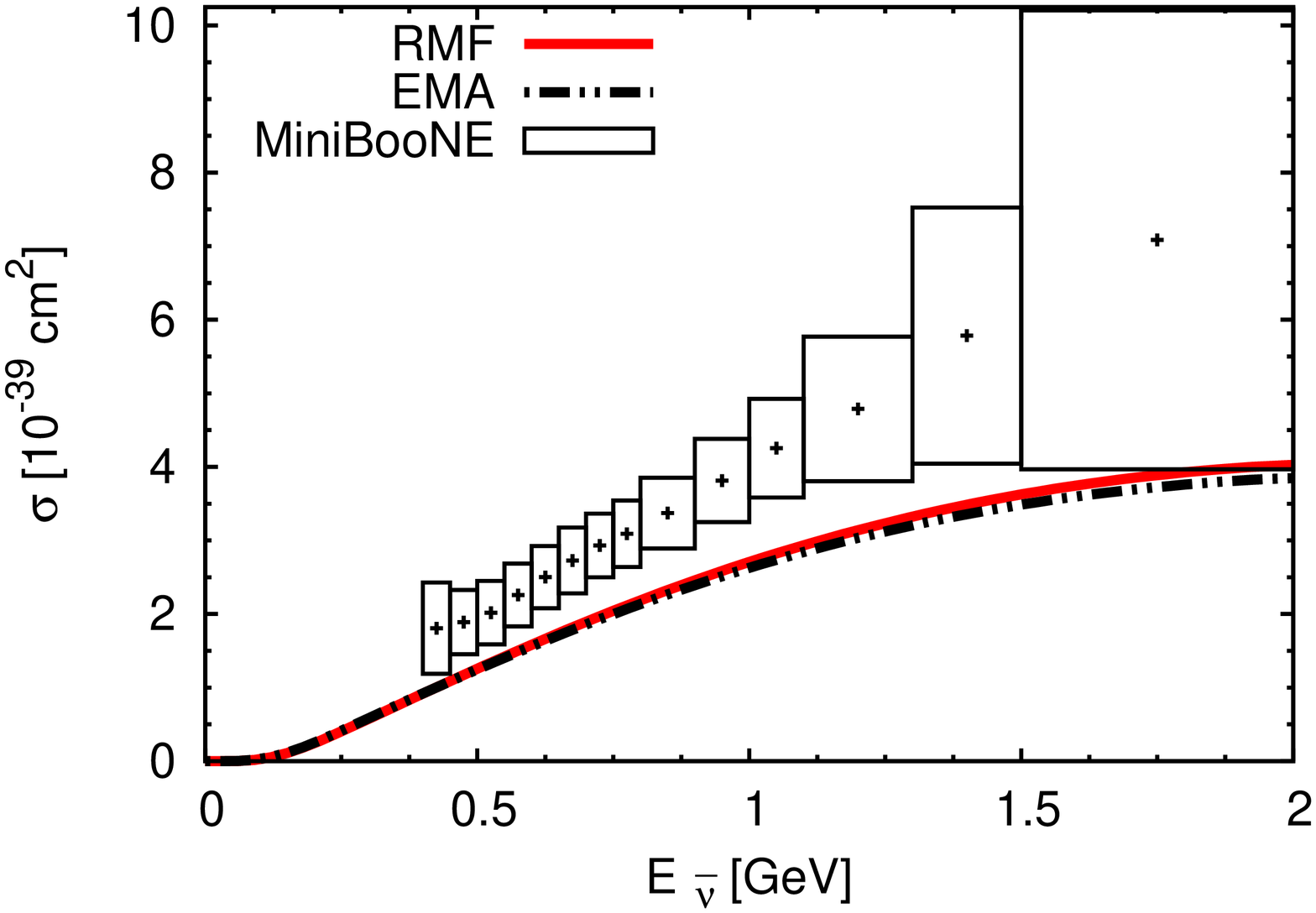}
\caption{(Color online) The same as Fig.~\ref{fig05}, but for CCQE $\overline\nu_\mu$-$^{12}$C scattering. Results are obtained within RMF model and EMA approach. The data are from \cite{AguilarArevalo:2010zc}.}\label{fig06}
\end{figure*}

With this guidance, it is easy to understand the results illustrated in Figs.~\ref{fig00}--\ref{fig06}. For instance in Fig.~\ref{fig00} we show the differential cross section per target nucleon for the (anti)neutrino CCQE process on $^{12}$C as a function of muon kinetic energy $T_\mu$. The incident (anti)neutrino energy is assumed to be $1$~GeV. In the figure results are given for the transverse ($T$), longitudinal ($L$), and axial-transverse ($|T'|$) contributions using the EMA (dashed lines, black) approach and the RMF (solid lines, red) model, respectively.

In Fig.~\ref{fig01} we present the flux-integrated double-differential cross section per target nucleon for the
$\nu_\mu$ CCQE process on $^{12}$C. We display the cross section, evaluated with the RMF model and EMA  approach, versus the $\mu^-$ kinetic energy $T_\mu$ for two bins of $\cos\theta_\mu$ (forward angles -- top panel and backward angles -- bottom panel of Fig.~\ref{fig01}). Also, in Fig.~\ref{fig01} are shown the separate contributions of
longitudinal ($\sigma_L$) and transverse ($\sigma_T$ and $\sigma_{T'}$) components calculated within the RMF and EMA
approaches. Here and in the following figures the results are compared to the MiniBooNE experimental
data~\cite{AguilarArevalo:2010zc, CCQE-anti-exp}. As shown, RMF and EMA results for the cross sections lie very close together, so the effects linked to the enhancement of the lower components due to the strong relativistic potentials are small. This conclusion also holds for antineutrino double-differential cross sections (Fig.~\ref{fig02}) as well as for differential and total unfolded integrated neutrino/antineutrino cross sections (Figs.~\ref{fig05} and~\ref{fig06}). Notice also the minor role played by the longitudinal component for angles in the range $0\leqq \theta_\mu \leqq 45$ degrees, this contribution being almost negligible for larger angles (as can be seen at backward angles -- bottom panel of Fig.~\ref{fig01}). On the contrary, as can be seen in Fig.~\ref{fig01}, most of the neutrino CCQE cross section comes from the pure transverse contribution given by the sum $\sigma_T+\sigma_{T'}$. The $T'$ contribution increases with the muon scattering angle $\theta_\mu$: its contribution at forward angles ($\cos\theta_\mu\sim 1$) is close to $L$ one, whereas at backward angles ($\cos\theta_\mu\sim -1$) it is almost equal to the $T$ contribution.

In Fig.~\ref{fig02} we present our predictions for the flux-averaged antineutrino CCQE cross sections corresponding to the MiniBooNE experiment~\cite{CCQE-anti-exp}. Here, in contrast with the neutrino case, the longitudinal contribution to the cross section plays a significant role, increasing its strength as the muon scattering angle $\theta_\mu$ goes up (as can be seen at backward angles -- bottom panel of Fig.~\ref{fig02}). This result can be understood from the destructive interference occurring between the two transverse responses, $T$ and $T'$. Note that in the case of antineutrinos the global transverse contribution to the cross section is given through the difference $\sigma_T-\sigma_{T'}$. On the contrary, neutrino reactions involve a constructive interference of both transverse responses. It is important to point out that $T$ and $T'$ contributions are much larger than $L$; however, they tend
to cancel for antineutrinos, hence explaining the relatively more significant role played by the longitudinal component in this case.

The difference between EMA and RMF results for $T$ and $T'$ is very similar. In the neutrino case we see in the total cross section mostly the same comparison of EMA to RMF as for the separate $T$ and $T'$ responses, namely about a few percent difference. However, for antineutrinos the effect of RMF versus EMA in $T$ and $T'$ responses, due to the change of sign, is cancelled to a large extent, at the same level as the $T$ and $T'$ responses are cancelled out causing the $L$ response to dominate the total cross section. Thus, for antineutrinos, for the cases where the total response is relatively small, there is no effect or difference between EMA and RMF results (Figs.~\ref{fig02} and \ref{fig04}).

In Fig.~\ref{fig03} (Fig.~\ref{fig04}) we present the flux-integrated double-differential cross section per target nucleon for the $\nu_\mu$ ($\overline{\nu}_\mu$) CCQE process on $^{12}$C for various bins of $\cos\theta_\mu$. As discussed above, RMF slightly exceeds EMA results for neutrino scattering due to the sum of $T$ and $T'$ contributions which for all angles are bigger than EMA ones. For  antineutrino CCQE process on $^{12}$C, results are almost identical within two approaches, only at large backward angles there are small differences. As can be seen from Figs.~\ref{fig03} and~\ref{fig04}, theoretical predictions clearly underestimate the experimental cross sections for neutrinos and antineutrinos. This result is consistent with the additional strength, not included in our model, that may come from two-body currents and multi-nucleon processes. While the RMF approach may account for some effects linked to two-body contributions, there would certainly be additional contributions beyond the IA lacking in this model.

In Fig.~\ref{fig05} (Fig.~\ref{fig06}) results are presented for the MiniBooNE flux-averaged CCQE $\nu_\mu$($\overline{\nu}_\mu$)-$^{12}$C differential cross section per nucleon as a function of the muon scattering angle (left-top panel, note that in order to compare with data the integration is performed over the muon kinetic energies $0.2~\text{GeV} < T_\mu < 2.0~\text{GeV}$), the muon kinetic energy (right-top) and the four-momentum transfer $Q^2$ (left-bottom). For completeness, we also show the total flux-unfolded integrated cross section per nucleon versus the neutrino energy (right-bottom). As in the previous figures, the use of the standard value for the axial mass within our model leads to results clearly below the data. However, the shape of the cross section is reproduced by the RMF model and EMA approach. Also, we note that the RMF model (which uses off-shell nucleon wave functions) yields larger transverse $T$ and $T'$ contributions than the EMA approach: this leads to an increase of the cross section within RMF for neutrino scattering compared to models without spinor distortions, whereas a cancellation of this effect is seen for antineutrino scattering.

\section{Conclusions}\label{sec4}

We have studied off-shell effects within a fully relativistic approach, the Relativistic Mean Field
model, which displays strong off-shell, non-factorizing behaviour. We can summarize our findings as follows:

1) Most theoretical approaches to CCQE neutrino scattering are based on factorization assumptions, or at least use on-shell, or almost on-shell, spinors to describe the nucleons. On the other hand the RMF model uses off-shell spinors, with strongly enhanced lower components. In this work we have studied the effect of this enhancement of the lower components for (anti)neutrino CCQE results. We have seen how these off-shell effects are visible, although being relatively small, in the total cross section for neutrino-nucleus scattering while, for MiniBooNE kinematics they are negligible for the antineutrino cross section. The effect of off-shell spinor distortion in the RMF cross-sections for neutrinos and antineutrinos can be compared with other ingredients considered in alternative approaches. For instance, in the case of \cite{Amaro:2010sd,Amaro:2011aa}, pionic MEC effects were studied and they were assumed to modify just the $T$ response, which is enhanced, whereas the $T'$ 2p2h excitations are suppressed and were accordingly neglected. Therefore those MEC effects are larger for antineutrinos than for neutrinos because the $T-T'$ cancellation is less severe. On the contrary, Martini \emph{et al.}~\cite{Martini2010} find a somewhat minor role of the 2p2h mechanisms for the antineutrino case. Finally, Nieves \emph{et al.}~\cite{Nieves:2011pp} get similar relative multi-nucleon contributions for neutrinos and antineutrinos. Although the conclusions about off-shell effects leading to spinor distortion considered here are based on a specific model, one has to recall that it is always the case that off-shell effects can only be studied within a model. However, from what we see here one can be reasonably confident that for MiniBooNE kinematics, dynamical  off-shellness leading to increased lower components in the nucleon spinors would be a rather small effect for neutrino-nucleus scattering, and fully negligible for the antineutrino-nucleus case. We have verified that the dynamical off-shell effects considered in this work affect the different contributions to the cross-section in a similar way as found in this work, for higher (anti-)neutrino energies, upto 100 GeV. This is due to the fact that for higher projectile energies, the cross-section is more and more forward peaked and then the momentum transfer is kept relatively small, no matter how large is the incoming lepton energy. The only thing to keep in mind is that the almost complete insensitivity to off-shell effects for the anti-neutrino case shown at MiniBooNE energies, depends on the cancellation of two contributions whose relative weight depends on the kinematics.
The cancellation of $T$ and $T'$ contributions breaks above $2$~GeV of incoming lepton energy. Actually, the $T'$ cross section becomes negligible at very high energies, so the sensitivity of the cross-sections to the enhancement of the lower components of the nucleon spinors would be similar for neutrino and antineutrino for neutrino energies of several GeV and above. In this work we have shown that the dependence of the neutrino and anti-neutrino cross-section to spinor distortion ambiguity is relatively small. This ambiguity would be hidden or remain unnoticed when using non-relativistic (in structure) models, but it should be kept in mind when trying to derive neutrino properties from experiments.

2) Neutrino MiniBooNE cross sections cannot be empirically fitted within several IA approaches: RFG, realistic Spectral Function approach, Super-Scaling-Approximation,  and RMF. An {\it ad hoc} enhancement of the axial mass, or what is the same, enhanced contribution from the axial term is needed for these models to explain the data. We have shown that this remains the case for antineutrino scattering.


\section*{Acknowledgements}
This work was partially supported by Spanish DGI and FEDER funds (FIS2011-28738-C02-01, FPA2010-17142), by the Junta
de Andalucia, by the Spanish Consolider-Ingenio 2000 program CPAN (CSD2007-00042), by the Campus of Excellence
International (CEI) of Moncloa project (Madrid) and Andalucia Tech, by the Istituto Nazionale di Fisica Nucleare under Contract MB31, by the INFN-MICINN collaboration agreement (AIC-D-2011-0704), as well as by the Bulgarian National Science Fund under contracts No. DO-02-285 and DID-02/16-17.12.2009. M.V.I. is grateful for the warm hospitality given by the UCM and for financial support during his stay there from the SiNuRSE action within the ENSAR european project.
R.G.J. acknowledges support from the Ministerio de Educaci{\'o}n (Spain) and T.W.D acknowledges support from the US Department of Energy under cooperative agreement DE-FC02-94ER40818.

\newpage

\end{document}